\documentclass[aps,prl,twocolumn,superscriptaddress]{revtex4-2}
\usepackage[utf8]{inputenc}
\usepackage{graphicx}
\usepackage{physics}
\usepackage{hyperref}
\usepackage[dvipsnames]{xcolor}
\usepackage{lipsum}

\begin{document}


\title{On the temperature dependence of the optical band gap in the material system of lithium niobate and lithium tantalate}


\author{Maximilian~Henneke}
\affiliation{Department of Physics, Paderborn University, Warburger Straße 100, 33098 Paderborn, Germany}

\author{Michael~Rüsing}
\affiliation{Department of Physics, Paderborn University, Warburger Straße 100, 33098 Paderborn, Germany}
\affiliation{Institute for Photonic Quantum Systems (PhoQS), Paderborn University, Warburger Straße 100, 33098 Paderborn, Germany}

\author{Nina~A.~Lange}
\affiliation{Department of Physics, Paderborn University, Warburger Straße 100, 33098 Paderborn, Germany}
\affiliation{Institute for Photonic Quantum Systems (PhoQS), Paderborn University, Warburger Straße 100, 33098 Paderborn, Germany}

\author{Timon Schapeler}
\affiliation{Department of Physics, Paderborn University, Warburger Straße 100, 33098 Paderborn, Germany}
\affiliation{Institute for Photonic Quantum Systems (PhoQS), Paderborn University, Warburger Straße 100, 33098 Paderborn, Germany}

\author{Noah Spiegelberg}
\affiliation{Department of Physics, Paderborn University, Warburger Straße 100, 33098 Paderborn, Germany}

\author{Ernst-Lukas Kuhlmann}
\affiliation{Department of Physics, Paderborn University, Warburger Straße 100, 33098 Paderborn, Germany}
\affiliation{Institute for Photonic Quantum Systems (PhoQS), Paderborn University, Warburger Straße 100, 33098 Paderborn, Germany}

\author{Elke Beyreuther}
\affiliation{Institute of Applied Physics, Technische Universität Dresden, Nöthnitzer Straße 61, 01187 Dresden, Germany}

\author{Philipp~Mues}
\affiliation{Institute for Photonic Quantum Systems (PhoQS), Paderborn University, Warburger Straße 100, 33098 Paderborn, Germany}

\author{Ludmila Eisner}
\affiliation{Institute for Energy Research and Physical Technologies,  Clausthal University of Technology, Am Stollen 19 B, Goslar, 38640, Germany}

\author{Lukas M. Eng}
\affiliation{Institute of Applied Physics, Technische Universität Dresden, Nöthnitzer Straße 61, 01187 Dresden, Germany}
\affiliation{ctd.qmat: Dresden-Würzburg Cluster of Excellence—EXC 2147, Technische Universität Dresden, 01062 Dresden, Germany}

\author{Laura~Padberg}
\affiliation{Department of Physics, Paderborn University, Warburger Straße 100, 33098 Paderborn, Germany}
\affiliation{Institute for Photonic Quantum Systems (PhoQS), Paderborn University, Warburger Straße 100, 33098 Paderborn, Germany}

\author{Donat J. As}
\affiliation{Department of Physics, Paderborn University, Warburger Straße 100, 33098 Paderborn, Germany}

\author{Klaus-Dieter Becker}
\affiliation{Institute for Physical and Theoretical Chemistry, Technische Universität Braunschweig, Gaußstraße 17, 38106 Braunschweig, Germany}

\author{Tim~J.~Bartley}
\affiliation{Department of Physics, Paderborn University, Warburger Straße 100, 33098 Paderborn, Germany}
\affiliation{Institute for Photonic Quantum Systems (PhoQS), Paderborn University, Warburger Straße 100, 33098 Paderborn, Germany}

\author{Christine~Silberhorn}
\affiliation{Department of Physics, Paderborn University, Warburger Straße 100, 33098 Paderborn, Germany}
\affiliation{Institute for Photonic Quantum Systems (PhoQS), Paderborn University, Warburger Straße 100, 33098 Paderborn, Germany}

\author{Christof~Eigner}
\affiliation{Institute for Photonic Quantum Systems (PhoQS), Paderborn University, Warburger Straße 100, 33098 Paderborn, Germany}


\date{\today}

\begin{abstract}
Lithium niobate and lithium tantalate see widespread use in optics and electronics, and are increasingly used for cryogenic applications. Despite their broad deployment, their optical band gap and its relation to the crystal stoichiometry are not well characterised as a function of temperature. In this work, we study the optical absorption properties of congruent, stoichiometric, MgO-doped and Er-doped lithium niobate as well as congruent lithium tantalate across the temperature range between 7~K and 1000~K by means of optical transmission spectroscopy. Our results demonstrate that the difference of the optical band gap typically observed at room temperature between different stoichiometries is not primarily attributable to the intrinsic electronic structure, but rather to different electron-phonon couplings and the average phonon energies. Additionally, we exemplarily study the temperature shift of the 523 nm absorption line in Er-doped lithium niobate due to the increased interest in optically active dopants. To facilitate future analyses, we present the open-source software suite PhoQS-Treat (Tauc Regression Edge Analysis Tool), which enables automated Tauc regressions alongside additional analytical capabilities. This work advances the development of high-performance lithium niobate-based devices.
\end{abstract}


\maketitle

\section{Introduction}

Lithium niobate (LNO, LiNbO$_3$) and lithium tantalate (LTO, LiTaO$_3$) see widespread use in research and industry due to their large piezoeletric, pyroelectric, electro-optic, as well as second-order nonlinear optical properties, which are combined with a large transparency window and the possibility for ferroelectric domain engineering \cite{Yue2003,weigel2023,sanchezdena2020,Mackwitz2016,Boes2023,Qi2020,Schroeder2012,Rogers2026,Lin2026,Qi2020}. Both materials can be fabricated in a scalable manner by the Czochralski-technique and are commercially available in form of large wafers of up to 200~mm diameter in both bulk or thin-film form. Consequently, the LiNbO$_3$  family plays a vital role in a diverse spectrum of experimental and commercial applications ranging from radio-frequency filters based on the surface and bulk acoustic effect, electro-optical modulators for high speed data communication, nonlinear and quantum optics, to more experimental applications like rewritable electronics based on ferroelectric domain walls \cite{cai2019,wang2018,xia2024,kießler2025,zhang2017,Boes2023,Qi2020,Schroeder2012,Rogers2026,Lin2026,Qi2020,Zahn2024,Ratzenberger2024}. More recently, also mixed crystals of LiNbO$_3$ and LiTaO$_3$ receive increasing interest, as they combine unique properties like the thermal stability of LiTaO$_3$ with the large piezoelectric response of LiNbO$_3$ \cite{becker2024,Sauerwein2026,Zabelina2023,Bashir2023}. One particular field of interest over the last years are cryogenic applications, which are central for quantum applications, e.g. as quantum interconnects, single-photon sources or combined with cryogenic single-photon detectors \cite{Lange2022cryogenic,lange2025widely,lange2023degenerate,Lee2025,Cheng2025}.

The reproducible fabrication of any devices operating over wide temperature ranges requires a thorough understanding of the material properties as well as their changes as a function of temperature. In this regard, the optical band gap represents a relatively easy to access and key fundamental parameter, which is directly related to the energy band structure. The electronic states govern properties like transmittance or the refractive index, which are key for (nonlinear) optical applications. Consequently, measurements of the optical band gap in the lithium niobate family are a standard characterisation method. For example, the optical band gap at room temperature is very sensitive to small changes in the lithium stoichiometry, which impacts refractive index, optical damage threshold and ferroelectric domain inversion, among other material properties \cite{kovacs1997, schlarb1994, abarkan2008}.

Furthermore, many ab-initio theoretical methods typically calculate materials properties only at 0~K \cite{Bernhardt2024,Veithen2002,Yang2013}, while accompanying experimental studies are performed at room temperature. In this regard, low-temperature measurements also provide valuable bench mark tests for theoretical studies.

While measurements of the optical band gap in LNO or LTO are reported regularly at room temperature (e.g. Refs. \cite{redfield1974,becker2024, bhatt2011}), very few studies have investigated the low temperature shift of the optical band gap (e. g. Ref. \cite{zanatta2022} down to $\sim$ 80 K or Ref. \cite{Bernhardt2024} down to 30 K). However, the low temperature studies in particular enable insight into the intrinsic (optical) band gap in the absence of phonon influences, which is a long standing challenge in lithium niobate \cite{thierfelder2010}. Therefore, in this study we investigate and compare the optical band gap of different, common stoichiometries of lithium niobate (congruent, stoichiometric, 5 mol-\% MgO-doped) and lithium tantalate (congruent) as a function of temperature from 300~K up to 1000~K and down to 7~K. The data is analysed in terms of the model by O'Donnell and Chen which allows to disentangle the impact of the phonon energies and electron-phonon coupling from the intrinsic optical band gap \cite{odonnell1991}. Additionally, we present the thermal behaviour of an Erbium absorption line down to 7~K. In this regard, LiNbO$_3$ doped with Er is studied because it is a potential candidate for quantum memories \cite{zhou2023}. Here, while the absorption spectra at room temperature are well known, the potential shift of the absorption lines is a key parameter when designing quantum memories in the future. To aid such and similar investigations, we further present the open source software suite PhoQS-Treat (Tauc Regression Edge Analysis Tool), which enables the direct analysis of optical transmission spectra datasets.

\section{Methods and Materials}

\begin{figure*}[!htb]
    \centering
    \includegraphics[width=1\linewidth]{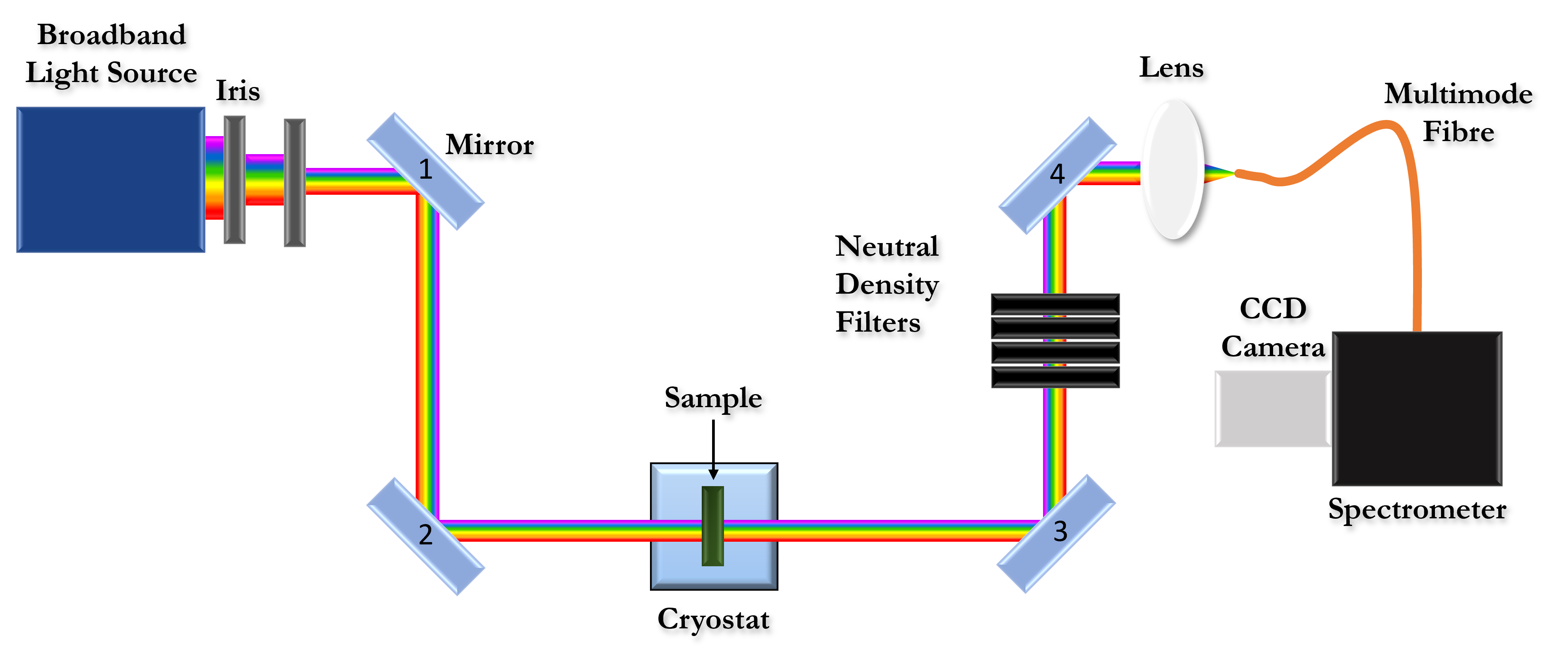}
    \caption{Schematic sketch of the setup used for the acquisition of transmission spectra. The operation range of the light source is 190 nm to 2500 nm. Al-coated mirrors were used because most measurements were carried out in the UV range.}
    \label{fig:setup}
\end{figure*}

LiNbO$_3$ and LiTaO$_3$ are isostructural crystals with similar properties. Both crystals can be grown as large single crystals from melts of LiO$_2$ and Nb$_2$O$_5$ or Ta$_2$O$_5$, respectively, using the Czochralski method. Crystals obtained through this process usually feature the congruent composition, which means a lower than ideal lithium content \cite{Carruthers1971,Kim2001b}. This lithium deficit is in the range of 1.0 to 1.5\% for LTO and LNO, respectively. This deficiency leads to rather high ratio of intrinsic defects, commonly Li-vacancies and Nb$_{\textnormal{Li}}$-antisites or Ta$_{\textnormal{Li}}$-antisites, i. e. niobium/tantalum ions on lithium lattice sites, which are required for charge compensation \cite{Gopalan2007,Dömer2024}. Although the Li deficiency is the main reason for various intrinsic defects leading to increased ferroelectric coercive fields or lowered optical-damage thresholds, congruent crystals are most commonly used in research or industry. For higher optical powers, e.g. in nonlinear frequency convertors, either (near-)stoichiometric, i.e. crystals with a close to 1:1 ratio of Li:Nb or Li:Ta, or Mg-doped crystals are used \cite{bordui1993,suizu2008}. In this regard, (near-)stoichiometric crystals do not have intrinsic defects caused by the lithium deficiency. However, stoichiometric crystals require a more elaborated and, hence, expensive fabrication \cite{Dravecz2006,Lengyel2015}. These can be obtained either during growth by controlling the stoichiometry in the melt \cite{kitamura1992}, or by post-fabrication treatment, e.g. vapour transport equilibration \cite{bhatt2012}. In contrast, Mg-doped crystals can be obtained with a much simpler growth scheme by adding MgO to the congruent melt \cite{tan1994}. In this process, Mg-ions displace Nb$_{\textnormal{Li}}$-antisites, which are mainly responsible for many of the detrimental effects, which requires a certain doping threshold needs to be reached \cite{polgar1986,Bocchini2025}. Here, typical doping concentrations in the range of 5 mol-\% are used for LNO to compensate for the intrinsic defects. Hence, Mg-doped crystals are usually very commonly available as well. 

Therefore, to gain insight into the most common stoichiometries we study single crystals of z-cut orientation of 1) congruent, 2) (near-)stoichiometric, 3) 5 mol-\% Mg-doped and 4) Er-doped lithium niobate, as well as 5) congruent lithium tantalate. All samples have been grown in their respective stoichiometry and doping by the Czochralski method. The congruent and Mg-doped LNO wafers were obtained from G\&H Crystals, while the congruent LTO samples were sourced commercially from The Roditi International Corporation Ltd. In contrast, the stoichiometric and Er-doped lithium niobate crystals were grown at the Wigner Institute, Budapest. For reference and to test the software suite, zinc-blende type silicon carbide (3C-SiC) and hydrothermally-grown potassium titanyl phosphate (hKTP) were measured as well. In contrast to LNO or LTO, 3C-SiC is a non-polar and nonferroelectric crystal, which is expected to have a much lower temperature-dependent band gap shift.

The optical band gap is measured by means of absorption spectroscopy, where a broadband light source is directed onto the sample and the transmitted light is spectrally analysed. Importantly, this measurement does not yield the electronic band gap, i.e. the energy difference of the lowest conduction band level and the highest valence band state, but usually a smaller band gap. This is well known for the LNO family and is primarily caused by rather strong excitonic effects \cite{thierfelder2010,riefer2013,Friedrich2017,Schmidt2022}, which results in optically active levels up to 700~meV below the electronic band gap. Hence, optical band gap measurements might only be directly compared to band gap determinations with other methods or theoretical calculations, which do not always consider excitonic effects.

\subsection{Experimental Setup}
The setup used for the room-temperature and low-temperature measurements is shown in Fig. \ref{fig:setup}. It consists of a broadband light source, a sample mount placed  within a cryostat and a spectrometer with attached CCD camera to achieve the spectral analysis. The components are discussed in detail below.\\

The broadband light source (operation range: 190 nm - 2500 nm) used is an Energetic EQ-99 LDLS able to reach a broadband optical power of 95 mW. UV-reflecting aluminium-coated mirrors were used as well as neutral density filters regulating the intensity of the light source in order to prevent detector saturation.  After the final mirror, the light is focused on a high-OH multimode fibre with a core diameter of 200 µm and a transmission down to 250 nm and guided to the Andor Kymera 193i spectrograph operating with a Czerny-Turner setup with an Andor iDus CCD camera attached. The CCD chip was cooled to -60 °C to reduce thermal noise. Its pixel size is 15 µm $\cdot$ 15 µm. A 50 mm $\cdot$ 50 mm large grating with 1200 $\mathrm{l\cdot mm^{-1}}$ and a blaze wavelength of 500 nm was used for all measurements, leading to a resolution limit of the spectrograph of 0.21 nm according to the data sheet.\\

The room temperature measurements were conducted with the corresponding Andor SOLIS software for individual measurements, while a Python script was used for the low temperature measurements. Here, the acquired spectra were stored with timestamps during the cooldown cycles. Using the latter method, data was acquired approximately every five seconds. The timestamps were later used to assign temperatures to the acquired spectra using the cryostat's log files. The cryostat was an Attocube attoDRY-system with optical windows.

\begin{figure*}
    \centering
    \includegraphics[width=1\linewidth]{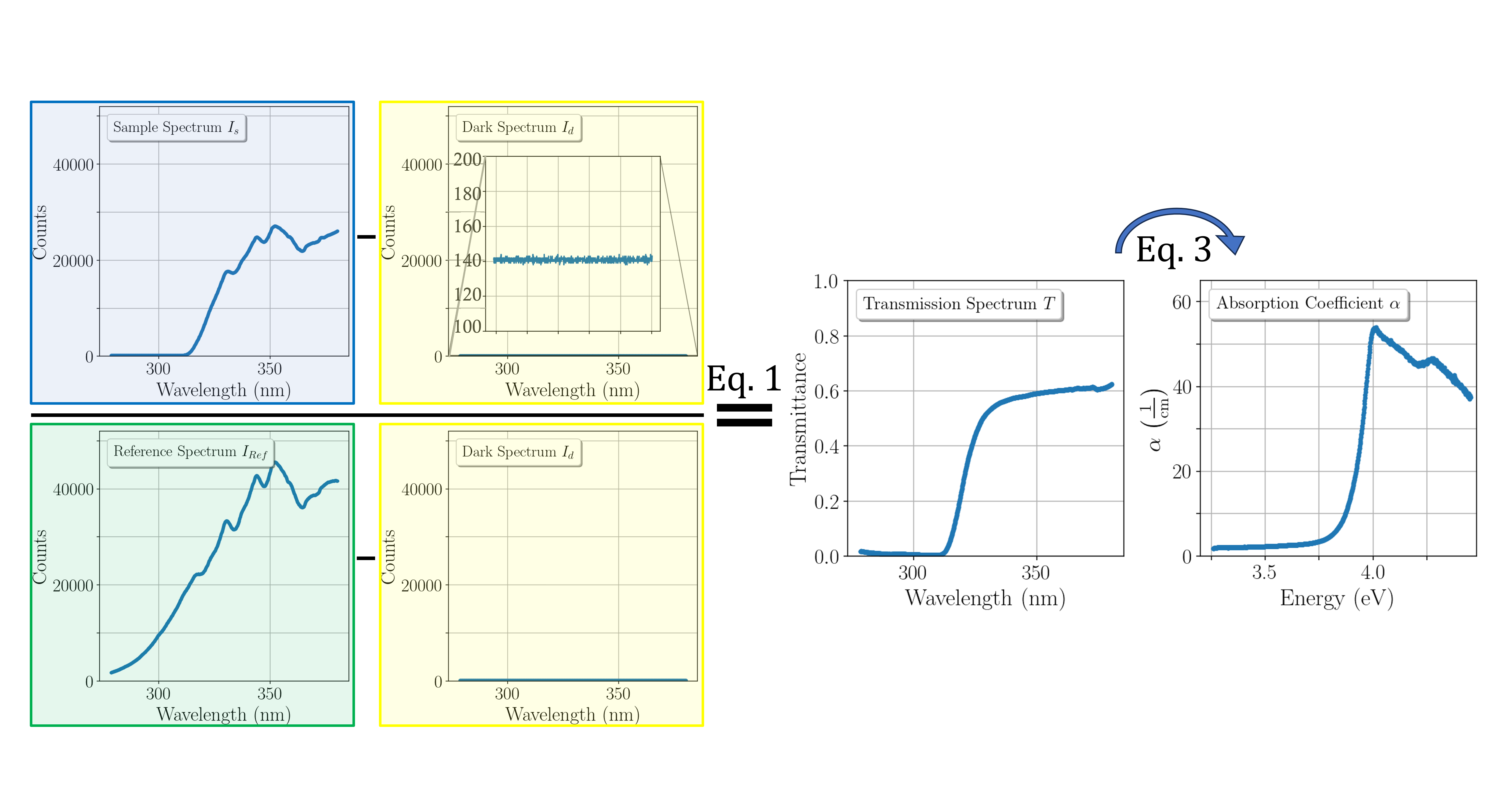}
    \caption{Illustration of the data processing using Eq. \ref{eq:transmittance} and Eq. \ref{eq:alpha} with the sample spectrum $I_s$ (blue), the reference spectrum $I_d$ (green) and the dark spectrum (yellow). The spectra belong to cLNO.}
    \label{fig:processing}
\end{figure*}

During a typical measurement run, at least three spectra are recorded to obtain a transmission spectrum $T$. These include a dark spectrum $I_d$ without any light source to characterise any detector or background noise, a reference spectrum $I_{Ref}$ which is a spectrum of the light source without any sample inserted in the path as well as the sample spectrum $I_s$. The calculation procedure is visualised in Fig.~\ref{fig:processing} and explained in the following section.

The high-temperature transmission spectra were measured using a Perkin-Elmer Lambda 900 double-beam photospectrometer modified for such experiments. As the focus of this work is primarily the fundamental absorption edge, the chosen wavelength range was 250~nm to 470~nm. The spectra were measured using a home-made heating chamber at temperatures ranging from room temperature to 727 °C. Further details concerning this setup can be found in Ref. \cite{becker2024}. It has to be noted that this spectrometer calculated the transmission spectrum after a measurement run internally. Therefore, the data processing and analysis of the high-temperature spectra started one step later with the calculation of the energy-dependent absorption coefficient $\alpha(E)$. Additionally, it has to be mentioned that the stoichiometric lithium niobate sample could not be investigated at high temperatures due to its dimensions.

\subsection{Data analysis procedure}

For each measurement, the spectral transmittance $T$ was obtained from the recorded sample spectrum $I_s$, reference spectrum $I_\mathrm{Ref}$ and dark spectrum $I_d$ according to
\begin{equation}
    \label{eq:transmittance}
    T = \frac{I_s-I_d}{I_\mathrm{Ref}-I_d}.
\end{equation}
The dark spectrum accounts for the detector background, while the reference spectrum represents the incident light intensity in the absence of the sample.

The measured transmittance is not determined by absorption alone. For a plane-parallel sample, the incident light is first partially reflected at the front surface, then attenuated inside the material according to the Lambert-Beer law,
\begin{equation}
    I(d) = I_0 \exp(-\alpha d),
\end{equation}

and finally partially reflected again at the rear surface. Here, $\alpha$ is the absorption coefficient and $d$ is the sample thickness. 

Since the physical information relevant for the band gap analysis is contained in $\alpha$, the absorption coefficient has to be extracted from the measured transmittance by separating absorption from interfacial reflection effects.

The absorption coefficient $\alpha$ can be obtained from the transmission data $T$ using an expression derived from an equation by MacFarlane and coworkers \cite{macfarlane1958} (see supplement for the derivation):

\begin{equation}
    \label{eq:alpha}
    \alpha = - \frac{1}{d}
    \ln\left(
    \frac{-(1-R)^2+\sqrt{(1-R)^4+4R^2T^2}}
    {2TR^2}
    \right).
\end{equation}
The reflectivity was estimated from the refractive index $n$ using the Fresnel expression for perpendicular incidence \cite{jiangou1992},
\begin{equation}
    \label{eq:reflectivity}
    R = \frac{(n-1)^2}{(n+1)^2}.
\end{equation}

The extracted absorption coefficient was used to determine optical band gaps by means of Tauc analysis. The Tauc relation is given by \cite{macyk2018}
\begin{equation}
    \label{eq:Taucmodel}
    (\alpha h\nu)^{1/\gamma} = B(h\nu-E_g),
\end{equation}

where $h$ is Planck's constant, $\nu$ is the photon frequency, $E_g$ is the optical band-gap energy, $B$ is a proportionality constant and $\gamma$ depends on the type of optical transition. For direct allowed transitions, $\gamma=0.5$, whereas for indirect transitions, $\gamma=2$. In this work, the analysis focuses on direct transitions; therefore, the band gap was determined from the linear region of the $\alpha^2$ representation. The Tauc regression was performed on $\alpha^2$ instead of $(\alpha h\nu)^2$, since the additional factor $h^2\nu^2$ only affects the scaling over the narrow fitting range used here. As the regression range is typically only a few meV wide, the resulting deviations are negligible; the inclusion of the factor is more important when large energy ranges are covered \cite{macfarlane1958}. A comparison using the full $(\alpha h\nu)^2$ representation is provided in the supplement. Both approaches can be found in the literature \cite{becker2024,bhatt2012}.

For the numerical analysis, the spectra were processed using the open-source software PhoQS-Treat programmed for this purpose. The software was used to calculate transmittance spectra with Eq.~\ref{eq:transmittance}, extract the absorption coefficient $\alpha$ using Eq.~\ref{eq:alpha}, and perform the Tauc regressions required for the band gap determination. For the band gap determination, the linear fit was performed in the vicinity of the absorption edge after applying a second order Savitzky-Golay filter with a window length of 80 x-axis units to the reference and sample spectra. A correlation threshold of 0.995 was used for all samples to identify sufficiently linear fitting regions. This threshold defines the minimum correlation coefficient of a linear test function required for accepting a segment as linear and is a user-adjustable parameter in the software; lowering the threshold allows less strictly linear regions to be included, whereas a higher value restricts the fit to regions with stronger linearity. The minimum fit-window lengths were set to 0.03 eV for the $\alpha^2$ representation and 0.05 eV for the $\sqrt{\alpha}$ representation. Further details on the implementation, regression algorithm, treatment of refractive indices and additional tools are provided in the supplement together with the software reference \cite{phoqstreat}.

For consistency, a standard thickness of $d=0.1$ cm was used for all samples. The sample thickness enters the calculation of $\alpha$ as a scaling parameter and therefore changes the absolute magnitude of the absorption coefficient. However, it does not shift the absorption edge and was not found to significantly affect the extracted band gap energy within the investigated range of thicknesses, as illustrated in the supplement.

The refractive indices entering Eq.~\ref{eq:reflectivity} were treated as constant within the evaluated spectral range. In principle, the refractive index depends on wavelength, temperature, composition and polarization and could therefore be described using suitable Sellmeier equations. However, such a treatment would require dispersion relations covering the stoichiometry, polarization, temperature range and spectral range of each sample. Test calculations using representative Sellmeier equations (\cite{zelmon1997} for cLNO, \cite{juvalta2006} for cLTO) showed that the resulting direct band gaps deviated by only some 10 meV from those obtained with constant refractive indices. Therefore, constant refractive indices were used in this work, namely $n=2.2$ for samples of the LNO/LTO family, $n=2.7$ for 3C-SiC and $n=1.8$ for hKTP \cite{schlarb1993,abedin1996,Shaffer1971,Kato2002}.\\
PhoQS-Treat also offers a tool for the identification of spectral features based on finding deviations larger than the spectrum’s uncertainty or, as an alternative, on finding the zero crossings in the derivative of the spectrum which are methods from Astropy's specutils package \cite{astropyI,astropyII,astropyIII}. Once a spectral line is identified, it is fitted with a Gaussian, a Lorentzian and a Voigt profile. Their respective parameters can be exported for further analysis. This feature will be used for the analysis of spectral features of the Er-doped crystal.

\section{Results}
\subsection{Room temperature band gaps}
\begin{figure*}
    \centering
    \includegraphics[width=1\linewidth]{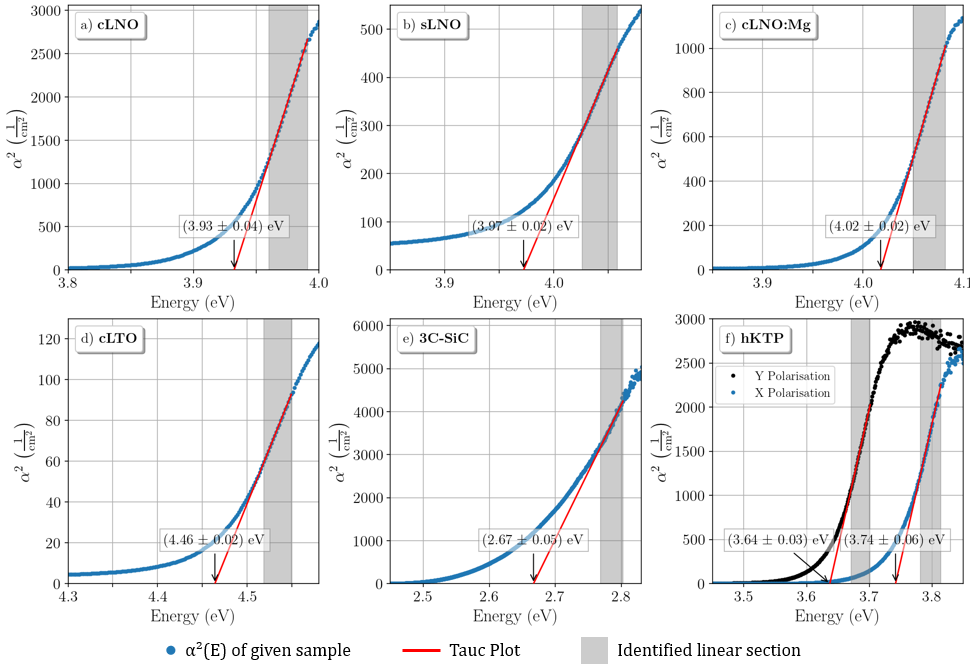}
    \caption{Plots of $\alpha^2(E)$ vs photon energy $E = h\nu$ for the determination of direct band gaps $E_g^d(300~\mathrm{K})$ via Tauc plots in a) cLNO, b) sLNO, c) cLNO:Mg, d) cLTO, e) 3C-SiC and f) hKTP at room temperature. For hKTP, the band gap was measured with x- and y-polarised light, respectively.}
    \label{fig:bandgaps}
\end{figure*}

Fig. \ref{fig:bandgaps} displays the plot of the squared absorption coefficient $\alpha^2(E)$ as a function of the photon energy $E$ for the determination of the direct band gap $E_g$ of the different samples at room temperature with Tauc plots. The linear data segment identified by PhoQS-Treat is highlighted with gray background.\\
Figs. \ref{fig:bandgaps}a)-d) show the results for the materials of the lithium niobate-tantalate family. The identified optical band gaps are consistent with known material properties from literature, such as the larger band gap of sLNO compared to cLNO \cite{kovacs1997} or that the optical band gap of the Mg doped congruent sample is closer to the stoichiometric sample than to the undoped, congruent one at room temperature (compare e. g. Refs. \cite{bhatt2012, polgar1986}). This can be explained with the incorporations of Mg$^{2+}$ ions on Nb antisites in the congruent crystal, leading effectively to a stoichiometric composition due to the compensation of the intrinsic defects \cite{kling2021}. Regarding the samples of the LNO/LTO family, the optical band gaps are in reasonable agreement with values reported in literature; some of the reported values are summarised in Table \ref{tab:bandgaps}.

The semiconductor 3C-SiC, see Fig.~\ref{fig:bandgaps}e), shows overall a similar picture to the LNO-LTO samples. Here, the material features a much higher band gap at around 2.67~eV compared to literature \cite{morkoc1994}.\\
The results for hKTP displayed in Fig.~~\ref{fig:bandgaps}f) show another interesting behaviour. In principle, hKTP is a wide band gap ferroelectric oxide crystal, which is comparable to LNO or LTO in its applications and transmission properties \cite{Wang2000,Cheng1993} with a room temperature optical band gap in the 3.5 to 4 eV range. However, in contrast to LNO or LTO, hKTP shows a very pronounced anisotropy in transmission, where the transmission edge depends on the light polarisation with respect to the crystal axis. This behaviour is also reproduced in our measurements demonstrating the sensitivity of the experimental setup and the software analysis. The determined optical band gaps are slightly higher than the values reported in literature. However, they still follow the trend that the band gap for x-polarised light is larger than the one for y-polarised light in KTP. Additionally, it has to be noted that Ref. \cite{hansson2000} defines the optical band gap as the wavelength corresponding to $\alpha=2$ cm$^{-1}$ and does not extract it via Tauc plots or similar methods.\\
Still, the investigations on hKTP and 3C-SiC justify the applied method as the results are comparable to literature.

\subsection{Band gaps at cryogenic temperatures}

The values obtained for the direct band gap at room temperature provide a starting point for the analysis of the behaviour of the band gaps at cryogenic temperatures. The temperature dependence of the band gap $E_g(T)$ can be described with the model developed by O'Donnell and Chen \cite{odonnell1991} which allows to disentangle the phononic influence on the band gap at absolute zero $E_g(0 \hspace{0.5mm}\mathrm{K})$ via the introduction of an electron-phonon coupling parameter $S$, also known as Huang-Rhys factor \cite{Yong_Zhang_2019, Huang1950}, and an average phonon energy $\langle\hbar\omega\rangle$:
\begin{equation}\label{eq:odonnellchen}
    E_g(T)=E_g(0\hspace{0.5mm}\mathrm{K})-S\langle\hbar\omega\rangle\Big[\coth\Big(\frac{\langle\hbar\omega\rangle}{2k_BT}\Big)-1\Big]
\end{equation}
In Eq. \ref{eq:odonnellchen}, $k_B$ denotes Boltzmann's constant. With this model, it is possible to quantify the shift of the direct absorption edge with temperature.

\begin{table*}
    \centering
    \begin{tabular}{cccccccc}
        Material & $E_g^d(300\hspace{1mm}\mathrm{K})$ [eV] & $E_g^i(300\hspace{1mm}\mathrm{K})$ [eV] &$ E_g(0\hspace{1mm} \mathrm{K})$ [eV] & $S$ & $S^+$ & $\langle\hbar\omega\rangle$ [meV] & Ref.\\
         \hline
        cLNO & $3.93\pm0.04$ & $3.82\pm0.02$ & $4.042\pm0.001$ & $7.6\pm0.3$ & $6.6\pm0.1$ & $46.0\pm1.7$ & This work \\
         & 4.03 & 3.8 & $4.2\pm0.1$ &  & 7.6 & $25\leq\langle\hbar\omega\rangle\leq50$ & \cite{becker2024}\\
         &  & $3.77 \pm 0.05$ & 3.96 & 7.775 & & 40 & \cite{zanatta2022} \\
         & 3.93 & 3.74 &  & & & &\cite{bhatt2012}\\
         & 3.91 &  &  & & & &\cite{joshi2022}\\
         \hline
        sLNO & $3.97\pm0.2$ & $3.78\pm0.01$ & $4.026\pm0.001$ & $5.5\pm0.6$ & & $52.4\pm4.1$ & This work\\
        & 4.10 & 3.91 &  & & & &\cite{bhatt2012}\\
        & 4.01 & 3.84 &  & & & &\cite{bhatt2011}\\
        \hline
        cLNO:Mg & $4.02\pm0.02$ & $3.90\pm0.02$ & $4.227\pm0.002$ & $9.3\pm0.4$ & $8.4\pm0.2$ & $40.7\pm1.9$ & This work\\
        & 3.99 &  & & & & & \cite{polgar1986}\\
        \hline
        cLTO & $4.46\pm0.02$ & $4.29\pm0.02$ & $4.6250\pm0.0007$ & $5.6\pm0.2$ & $7.5\pm0.1$ & $27.1\pm1.7$ & This work\\
         & 4.65 & 4.4 & $4.85\pm0.1$ & & 8.5 & $25\leq\langle\hbar\omega\rangle\leq50$ & \cite{becker2024}\\
         & 4.48 &  & & & & & \cite{kim2001}\\
         & 4.51 &  & & & & & \cite{hatano2004}\\
        \hline
        3C-SiC & $2.67\pm0.05$ & $2.41\pm0.03$ & $2.6986\pm0.0003$ & $1.3\pm0.2$ & & $36.1\pm5.2$ & This work\\
        & 2.2 &  & & & & & \cite{morkoc1994}\\
        \hline
        hKTP (X) & $3.74\pm0.06$ & $3.60\pm0.01$ & & & &  & This work\\
         & 3.52 &  & & & & & \cite{hansson2000}*\\
         \hline
        hKTP (Y) & $3.64\pm0.03$ & $3.51\pm0.01$ & & & &  & This work\\
         & 3.45 &  & & & & & \cite{hansson2000}*\\
    \end{tabular}
    \caption{Collection of the parameters direct and indirect band gap at room temperature $E_g^d,~E_g^i$, band gap at 0 K $E_g(0\hspace{1mm}\mathrm{K})$, electron-phonon coupling parameter $S$ obtained from cryogenic temperature data and $S^+$ obtained from elevated temperature data as well as average phonon energy $\langle\hbar\omega\rangle$ for the investigated materials (hKTP was studied at room temperature only).\\ *Ref. \cite{hansson2000} defines the cutoff wavelength (corresponds to direct band gap) as the x value corresponding to $\alpha=2\hspace{1mm} \mathrm{cm}^{-1}$.}
    \label{tab:bandgaps}
\end{table*}

\begin{figure*}
    \centering
    \includegraphics[width=1\linewidth]{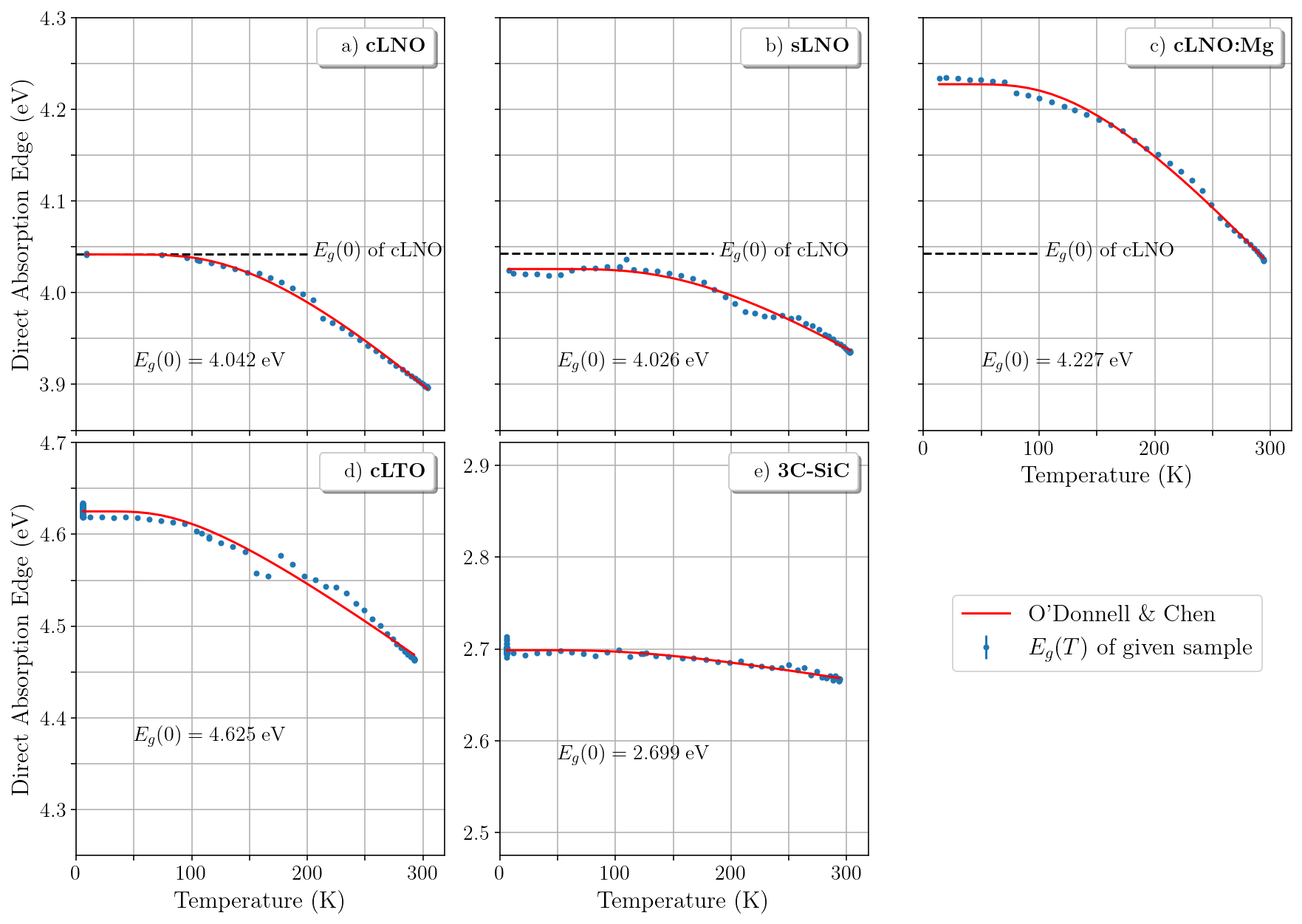}
    \caption{Plots of the direct absorption edge $E_g(T)$ over a temperature range from approx. 7 K to room temperature for a) cLNO, b) sLNO, c) cLNO:Mg, d) cLTO, and e) 3C-SiC fitted with the model of O'Donnell and Chen \cite{odonnell1991}. The band gap at absolute zero of each sample is also shown.}
    \label{fig:bandgaps_odonnell_chen}
\end{figure*}

Fig. \ref{fig:bandgaps_odonnell_chen} displays the direct absorption edges of a) cLNO, b) sLNO, c) cLNO:Mg, d) cLTO, e) 3C-SiC, and f) hKTP as a function of temperature. All graphs show the typical behaviour of an increasing band gap with decreasing temperature, which approaches a constant value at a certain temperature which is directly related to the electron-phonon coupling parameter $S$ and the average phonon energy $\langle\hbar\omega\rangle$.\\
To determine the values, the curves are fitted with the above-mentioned model (Eq. \ref{eq:odonnellchen}). The fit parameters as well as the  direct and indirect band gaps of all samples are collected in Table \ref{tab:bandgaps}.
The uncertainties were determined in two different ways: The uncertainty stated for the room temperature band gaps $E_g^d(300\hspace{1mm}\mathrm{K})$ is derived from the mathematical uncertainty of the regression procedure of one single measurement. Here the values are on the order of 0.05~eV. However, this mathematical procedure might overestimate the uncertainty in the measurements. Therefore, to estimate the uncertainty we computed the average and standard deviation based on 1000 band gaps of the cLTO sample determined from spectra measured at the same temperature (approximately 5.5 K), amounting to a variation of $\sigma(E_{g,cLTO})=0.0008$ eV, which is much smaller than the mathematical uncertainty for each fit. Therefore, this value was used as a further estimate for each measurement point in Fig.~\ref{fig:bandgaps_odonnell_chen}. Consequently, the uncertainties indicated in the Table~\ref{tab:bandgaps} for the fit parameters $ E_g(0\hspace{1mm} \mathrm{K})$, $S$, and $\langle\hbar\omega\rangle$ are a result of the statistical uncertainties of the fitting procedure of the O'Donnell\&Chen model.\\

These uncertainties are small compared to the observed shifts. Here, it can be noticed that in particular the LNO and LTO samples show some pronounced jumps in the datasets, which are larger than these uncertainties. These jumps are probably related to pyroelectric charge built-up and associated discharging, which has been observed before in opto-electronic investigations of LNO cooled down to cryogenic temperatures \cite{Thiele2024b,Lange2026}. While the slow cooling rate and the broadband super-band gap illumination during our experiment, which creates free charge carriers, can reduce the build-up of pyroelectric charges, it cannot be fully suppressed. This is due to the very small electrical conductivity of LNO at low temperatures \cite{Zahn2024}. This effect also leads to a hysteresis behaviour for some samples as shown in the supplement, which during heat up do not follow the same path. Therefore, we have analysed only the cooling-cycles, because they generally show a more consistent behaviour.\\

Regarding the identified parameters $E_g(0~\mathrm{K)}$, $S$ and $\langle\hbar\omega\rangle$ of the O'Donnell\&Chen model, it has to be mentioned first that the band gaps of cLNO and sLNO at absolute zero deviate only by about 0.4\% for the investigated samples. The Huang-Rhys factor $S$, however, is higher in the cLNO sample, indicating that the band gap difference between the congruent and stoichiometric composition observed at room temperature is actually not a result of deviations in the intrinsic electronic and crystal structure, but rather of different coupling strengths between electrons and phonons. This might arise from the defect structure, because the increased defect density in congruent lithium niobate allows for better coupling of the phonons to the crystal lattice. The improved lattice quality could also be an explanation for the higher average phonon energy of sLNO, as defects lead to a smearing of the phonon density-of-states.\\
In the Mg-doped sample, the band gap at absolute zero is slightly higher than in the two undoped LNO samples. This can be explained with the ionic radii of the Li$^+$ and the Mg$^{2+}$ ions, which are 0.76 Å and 0.64 Å, respectively \cite{tao2019, remko2006}. When LNO is doped with Mg above the doping threshold, the Mg$^{2+}$ ions are incorporated at the Li site \cite{schlarb1994}. As the Mg$^{2+}$ ion, which is a heterovalent dopant cation, is smaller than the Li$^+$ one, the lattice is compressed, which in turn leads to an increase of the band gap \cite{husin2019}. The $S$-parameter is even larger than for cLNO indicating a strong lattice disorder caused by the incorporation of Mg-ions \cite{Lengyel2007}; the average phonon energy is reduced even more by smearing of the phonon density-of-state.\\
In lithium tantalate, the band gap at absolute zero as well as at room temperature is higher compared to lithium niobate while the coupling parameter and the average phonon energy are lower. The lower $S$ parameter observed for room-temperature to cryogenic temperatures could be a result of the slightly improved lithium stoichiometry of a cLTO vs. a cLNO crystal. Usually, cLTO crystals are reported to feature a slightly lower lithium deficiency \cite{weigel2023}. The reduced average phonon energy is a direct result of the increased mass of the Ta-ion compared to the Nb-ion. \\
For comparison, the non-polar and non-ferroelectric 3C-SiC sample was measured. Here, a very low electron-phonon coupling constant is observed, which is 5 to 7 times smaller compared to that of the LNO and LTO samples. This highlights that the large thermal shifts of the band gap in LNO and LTO are related to their ferroelectric and polar nature.

\subsection{High Temperature Measurements}

\begin{figure*}[h]
    \centering
    \includegraphics[width=1\linewidth]{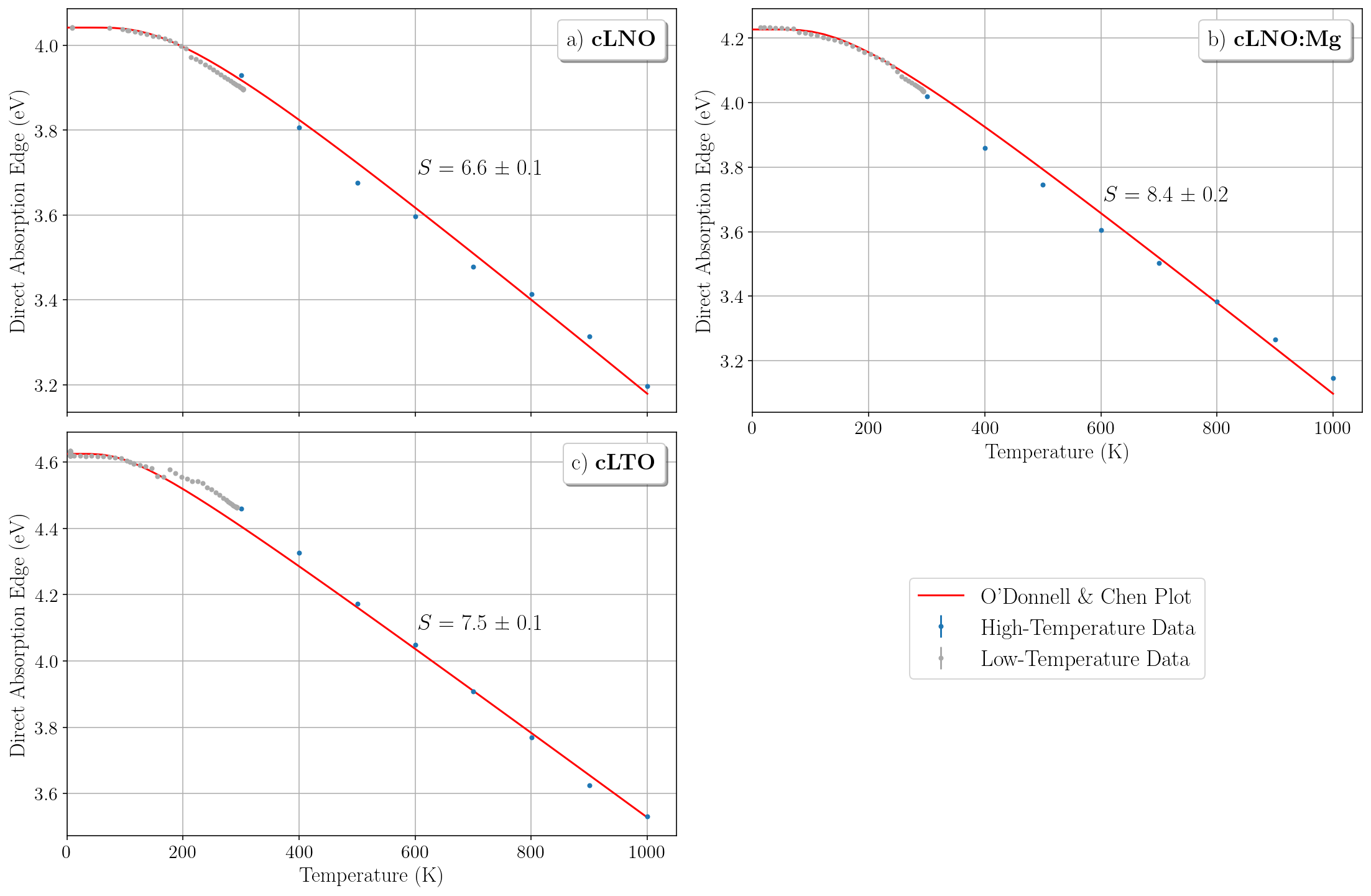}
    \caption{Temperature dependence of the optical band gap of a) cLNO, b) cLNO:Mg and c) cLTO in the temperature range between approx. 5 K and 1000 K as predicted by the model of O'Donnell \& Chen (Eq. \ref{eq:odonnellchen}). The two sets of experimentally acquired data as well as the electron-phonon coupling parameter $S^+$ obtained from the high temperature data are also shown.}
    \label{fig:high_t}
\end{figure*}

The combined datasets of high and low temperature measurements are shown in Fig. \ref{fig:high_t}. Here, only cLNO, cLNO:Mg and cLTO could be measured, as the sLNO crystal was to small to yield useful data in the high temperature setup.

While the measurements at cryogenic temperatures are well-suited to determine the band gap at absolute zero $E_g(0)$ and the average phonon energy $\langle\hbar\omega\rangle$, the Huang-Rhys factor $S$ is easier to derive from high temperature measurements for it describes the behaviour of the band gap after the tipping point defined by $\langle\hbar\omega\rangle$. Naturally, high temperature measurements are able to cover a broader temperature range in the linear regime of the band gap shift after this tipping point compared to measurements at cryogenic temperatures. Therefore, the high temperature data was used to improve the results for the $S$ parameter obtained from the low temperature data.

To this end, the previously determined parameters $E_g(0)$ and $\langle\hbar\omega\rangle$ were fixed for each of the samples, while the parameter $S$ was fitted to the combined dataset of high and low temperature data. This O'Donnell\&Chen model with reduced degrees of freedom was then fitted to the high temperature data. The resulting $S^+$ parameter is listed in Table \ref{tab:bandgaps}, while the final model of the temperature dependence of the optical band gaps of cLNO, cLNO:Mg and cLTO is depicted together with the acquired data in the two temperature regimes in Fig. \ref{fig:high_t}. 

It has to be mentioned that the high temperature data shown here is corrected by an offset for the data systematically showed a slightly higher band gap at room temperature for all data sets indicating slight calibration offsets between the setups. As we are mostly interested in the slope, the high temperature datasets were shifted according to the difference at room temperature. In addition, it was decided to use the uncertainty of the low temperature data for the high temperature data as well. This is justified by the fact that the resulting electron-phonon coupling parameters deviate only by a negligible amount when using larger uncertainties during the fit as shown in the supplement.\\

Comparing the Huang-Rhys factors $S$ obtained from the two temperature ranges, it is evident that the electron-phonon coupling constant of the low-temperature data does not show the whole picture, because the $S$ values of the high-temperature data differ significantly from their low-temperature counterparts. In fact, the order from lowest to highest coupling parameter is now cLNO - cLTO - cLNO:Mg in contrast to cLTO - cLNO - cLNO:Mg as indicated by the low-temperature data.\\

This can be explained with the number of intrinsic defects present in the crystal lattices: As mentioned above, the incorporation of Mg ions causes a high degree of lattice disorder as can be seen in Raman spectra \cite{Lengyel2007}, allowing a stronger coupling of phonons to the lattice and thus leading to a higher $S$ parameter of cLNO:Mg compared to the undoped crystal. Similarly, the lower electron-phonon coupling parameter of cLNO compared to cLTO could be a result of the presence of Ta ions at interstitial lattice sites in lithium tantalate, while there are no corresponding Nb ions at these sites in lithium niobate \cite{Dömer2024}.\\
In conclusion, it is not sufficient to consider only low-temperature band gaps to investigate the electron-phonon coupling parameter in the LiNbO$_3$/LiTaO$_3$ material system.

\subsection{Absorption Lines of cLNO:Er at Cryogenic Temperatures}

\begin{figure*}
    \centering
    \includegraphics[width=1\linewidth]{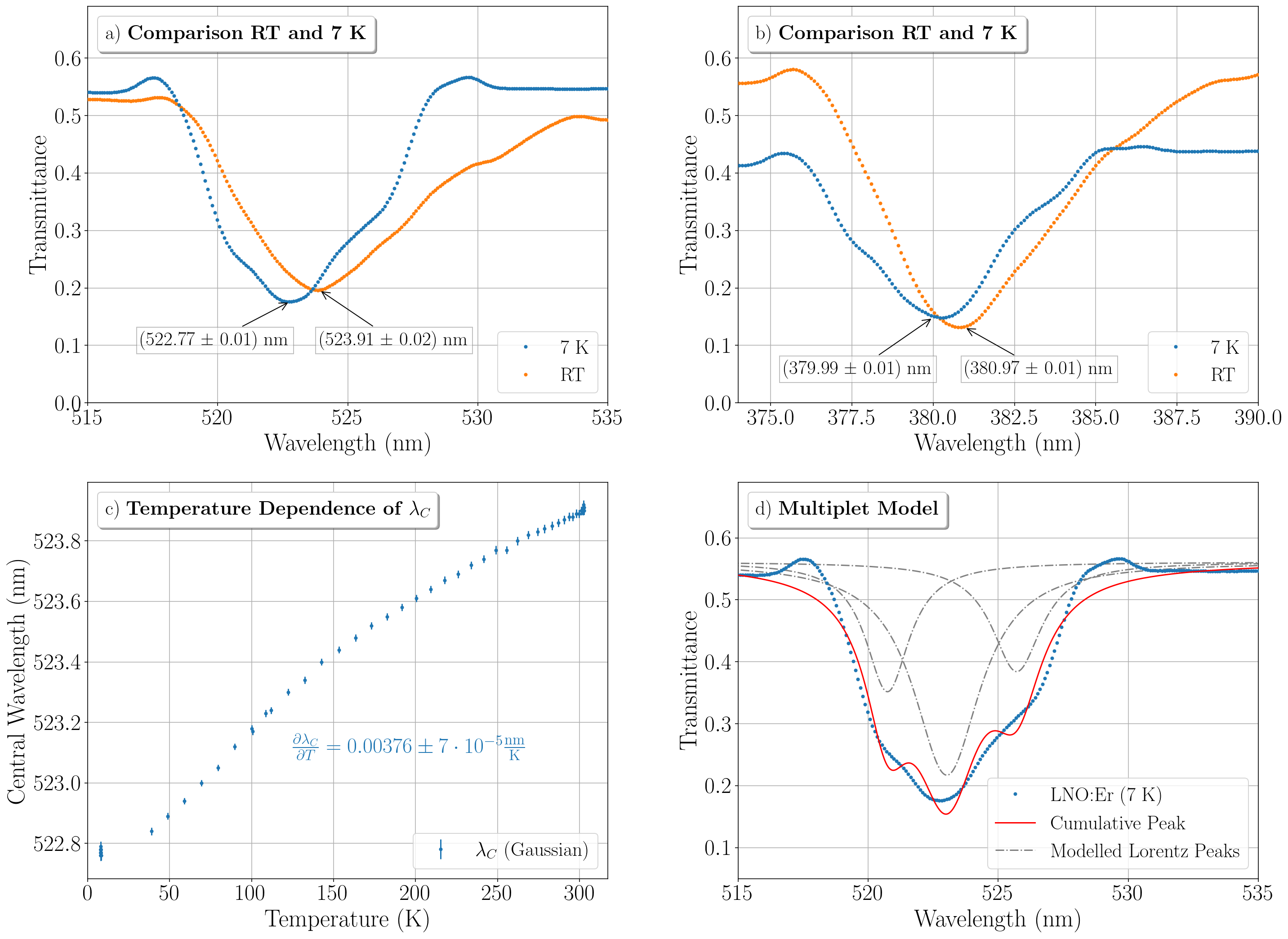}
    \caption{ Transmittance spectra of cLNO:Er at room temperature (orange) and at approximately 7 K (blue) of a) the 523 nm absorption line and of b) the 380 nm absorption line. The central wavelengths are highlighted. c) Temperature dependence of the central wavelength of the 523 nm absorption line. The given change of the central wavelength has been determined between room temperature and approx. 40 K. d) Modelling the absorption line with Lorentz peaks at 520.74 nm, 523.03 nm and 525.71 nm.}
    \label{fig:erbium}
\end{figure*}

Erbium is a technically widely used dopant due to its applications in the field of optical amplifiers and lasers in the telecom band. In recent years it sees renewed interest for incorporation in LNO and thin film lithium niobate to realise laser or quantum memories \cite{zhou2023,Wang2025,Wu2024,Dutta2023}. In this regard, cLNO:Er will also be applied in cryogenic environments, e.g. for increased coherence times. Therefore, possible shifts of optically active absorption lines as a function of temperature need to be characterised. Therefore, we exemplarily study the 523 nm absorption line of cLNO:Er, because it is close to wavelengths commonly used in lasers. In Fig. \ref{fig:erbium} a), the transmission spectrum of cLNO:Er between 515 nm and 535 nm is shown at 7 K and room temperature.

It is evident that the line displays a blue shift when the crystal is cooled down for the central wavelength of the absorption line has decreased by about 1 nm. This is also summarised in Table \ref{tab:erbium}, together with a second absorption line at about 380 nm also showing a blue shift for comparison, displayed in Fig. \ref{fig:erbium} b).

\begin{table*}
    \centering
    \begin{tabular}{ccc}
        Transition & Temperature [K] & Central wavelength [nm]\\
        \hline
        $^4\mathrm{I}_{15/2}\rightarrow^4\mathrm{G}_{11/2}$ & 300 & $380.97\pm0.01$\\
         & 7 & $379.99\pm0.01$\\
        $^4\mathrm{I}_{15/2}\rightarrow^2\mathrm{H}_{11/2}$ & 300 & $523.95\pm0.01$\\
         & 7 & $522.73\pm0.02$\\
    \end{tabular}
    \caption{Comparison of the central wavelengths of two of the absorption lines of cLNO:Er at room temperature and 7 K. The Er$^{3+}$ transitions were taken from \cite{sardar2003}.}
    \label{tab:erbium}
\end{table*}

To understand this shift better, the central wavelength $\lambda_C$ of the 523 nm line during cooldown from room temperature to approx. 7 ~K is shown in Fig. \ref{fig:erbium} c). As can be seen there, the central wavelength decreases rather linearly in the region between about 40 K and room temperature with a slope of $\pdv{\lambda_{C}}{T}=(0.00376\pm7\cdot10^{-5})$ $\frac{\mathrm{nm}}{\mathrm{K}}$ using a Gauss model for the identification of the central wavelength. A possible explanation for this shift might be that the crystal lattice contracts with decreasing temperature \cite{tarumi2012}. As the Er$^{3+}$ ions are not in the centre of their respective oxygen octahedra \cite{kling2021}, they are affected more and more by the potential of the contracting lattice when temperature decreases. This might cause a "re-centring" of the ion in the octahedron, leading to a shift of the central wavelength. As an analogy, one might consider a particle in a box, where the different energy levels are proportional to $L^{-2}$, with $L$ being the length of the box \cite{griffiths2018}.\\
Another possible explanation for the shift might be the Stark effect as LNO is ferroelectric and has therefore an internal electric field, which changes with temperature. For example, Skvortsov \textit{et al.} measured the Stark shifts in Chromium-doped LNO for the 520.6 nm line, which amounted to approximately -0.093 nm for fields of the order of 100 kV$\cdot$cm$^{-1}$ \cite{skvortsov1997}. The internal fields in cLNO account for about a fifth to a third of this value \cite{Gopalan2007}. However, these fields are amplified during cooldown due to emerging pyro-electric charges.\\ 
A first estimation of the amplification of the internal electric field can be made by modelling the sample as a plate capacitor. The pyro-electric coefficients of LNO are in the order of magnitude of $p\approx10^{-4}$ C/(Km$^2$) \cite{weigel2023}. For a plate capacitor, the electric field can be calculated according to 
\begin{equation}
E=\frac{Q}{\varepsilon_0\varepsilon_rA}=\frac{p\cdot\Delta T}{\varepsilon_0\varepsilon_r} .   
\end{equation}
With $\Delta T\approx300$ K and $\varepsilon_{r,LNO}\approx40$ \cite{michi2016}, the electric field amounts to $E\approx840$  kV$\cdot$cm$^{-1}$ under the assumption that the pyro-electric charges are not dissipating during the cooldown. As this is in the order of magnitude of the electric fields necessary to observe the above-mentioned Stark shift, less than 10\% of the total shift can be explained with this, assuming a similar behaviour to Chromium. Therefore, an intrinsic effect, like the re-centring, is likely the cause for the shift.\\
Another notable feature of Fig. \ref{fig:erbium}a) is the deviation of the absorption line from the expected single Gaussian profile, particularly at cryogenic temperatures. It seems that the line features a substructure. Therefore, the line was modelled with three individual Lorentzians positioned at 520.74 nm, 523.03 nm and 525.71 nm, respectively, as shown in Fig. \ref{fig:erbium} d). The cumulative peak resulting from the three modelled Lorentzians is on a reasonable level of consistency with the data, implying that the 523 nm absorption line is actually a multiplet. A possible reason might be Er-atoms incorporated at different lattice sites leading to different electronic environments, which results in distinct peaks. However, further investigations are necessary.\\

\section{Summary and Conclusion}
In this work, the temperature-dependent optical band gaps of lithium niobate (LNO) and lithium tantalate (LTO) were investigated over a temperature range from room temperature down to approximately 7 K. In addition, the open-source software suite PhoQS-TREAT (Tauc Regression Edge Analysis Tool) was developed to enable automated, reproducible, and user-independent determination of optical band gaps from transmission spectra using Tauc plot analysis.

The measurements demonstrate that the band gap difference commonly observed between congruent and stoichiometric LNO at room temperature is not an intrinsic property of their electronic structures. Instead, both materials exhibit nearly identical band gaps at 0 K, whereas the increasing deviation at elevated temperatures originates from differences in electron-phonon coupling and the corresponding average phonon energies. The stronger coupling observed in congruent LNO is attributed to its higher defect concentration, which enhances lattice interactions.

MgO doping was found to increase the optical band gap at 0 K, consistent with lattice compression resulting from the substitution of Li$^+$ by the smaller Mg$^{2+}$ ion. In Er-doped LNO, cooling to cryogenic temperatures resulted in a blue shift of the 523 nm absorption line by 1.14 nm. This behaviour is attributed to a temperature-induced re-centering of the Er$^{3+}$ ions within the oxygen octahedra as the crystal lattice contracts.

Furthermore, congruent LTO exhibits a consistently larger band gap and weaker electron-phonon coupling than congruent LNO over the entire investigated temperature range. A comparison with the non-polar semiconductor 3C-SiC, which displays substantially weaker electron-phonon coupling, indicates that the pronounced temperature dependence of the optical band gap in the LNO/LTO material family is closely related to their polar ferroelectric crystal structure.

Overall, this work provides new insights into the microscopic mechanisms governing the temperature dependence of the optical band gap in technologically relevant ferroelectric oxides. The presented methodology, together with the open-source software PhoQS-TREAT, establishes a reproducible framework for optical band gap determination, while the obtained material parameters provide valuable input for the design and optimisation of photonic and quantum technologies intended for operation under cryogenic conditions.

The presented results contribute to a deeper understanding of the interplay between lattice dynamics and electronic structure in ferroelectric oxides and provide a quantitative basis for modelling their optical properties over a wide temperature range. These findings are expected to facilitate the development of cryogenic photonic devices and rare-earth-based quantum technologies employing lithium niobate and lithium tantalate.

\begin{acknowledgments}
The DFG (TRR 142/3-2024 -- Project No.~231447078; FOR5044 -- Project No.~42670383) is gratefully acknowledged for financial support. We thank L. Kovács from the Wigner Institute, Budapest, for providing the stoichiometric and Er-doped samples.
\end{acknowledgments}

\section*{Data Availability Statement}
The datasets generated and analysed during the study are available on \url{https://doi.org/10.5281/zenodo.22011148}.

\begin{widetext}
    \title{Supplementary Material:\\
On the temperature dependence of the optical band gap in the material system of lithium niobate and lithium tantalate}
\author{Maximilian~Henneke}
\affiliation{Department of Physics, Paderborn University, Warburger Straße 100, 33098 Paderborn, Germany}

\author{Michael~Rüsing}
\affiliation{Department of Physics, Paderborn University, Warburger Straße 100, 33098 Paderborn, Germany}
\affiliation{Institute for Photonic Quantum Systems (PhoQS), Paderborn University, Warburger Straße 100, 33098 Paderborn, Germany}

\author{Nina~A.~Lange}
\affiliation{Department of Physics, Paderborn University, Warburger Straße 100, 33098 Paderborn, Germany}
\affiliation{Institute for Photonic Quantum Systems (PhoQS), Paderborn University, Warburger Straße 100, 33098 Paderborn, Germany}

\author{Timon Schapeler}
\affiliation{Department of Physics, Paderborn University, Warburger Straße 100, 33098 Paderborn, Germany}
\affiliation{Institute for Photonic Quantum Systems (PhoQS), Paderborn University, Warburger Straße 100, 33098 Paderborn, Germany}

\author{Noah Spiegelberg}
\affiliation{Department of Physics, Paderborn University, Warburger Straße 100, 33098 Paderborn, Germany}

\author{Ernst-Lukas Kuhlmann}
\affiliation{Department of Physics, Paderborn University, Warburger Straße 100, 33098 Paderborn, Germany}
\affiliation{Institute for Photonic Quantum Systems (PhoQS), Paderborn University, Warburger Straße 100, 33098 Paderborn, Germany}

\author{Elke Beyreuther}
\affiliation{Institute of Applied Physics, Technische Universität Dresden, Nöthnitzer Straße 61, 01187 Dresden, Germany}

\author{Philipp~Mues}
\affiliation{Institute for Photonic Quantum Systems (PhoQS), Paderborn University, Warburger Straße 100, 33098 Paderborn, Germany}

\author{Ludmila Eisner}
\affiliation{Institute for Energy Research and Physical Technologies,  Clausthal University of Technology, Am Stollen 19 B, Goslar, 38640, Germany}

\author{Lukas M. Eng}
\affiliation{Institute of Applied Physics, Technische Universität Dresden, Nöthnitzer Straße 61, 01187 Dresden, Germany}
\affiliation{ctd.qmat: Dresden-Würzburg Cluster of Excellence—EXC 2147, Technische Universität Dresden, 01062 Dresden, Germany}

\author{Laura~Padberg}
\affiliation{Department of Physics, Paderborn University, Warburger Straße 100, 33098 Paderborn, Germany}
\affiliation{Institute for Photonic Quantum Systems (PhoQS), Paderborn University, Warburger Straße 100, 33098 Paderborn, Germany}

\author{Donat J. As}
\affiliation{Department of Physics, Paderborn University, Warburger Straße 100, 33098 Paderborn, Germany}

\author{Klaus-Dieter Becker}
\affiliation{Institute for Physical and Theoretical Chemistry, Technische Universität Braunschweig, Gaußstraße 17, 38106 Braunschweig, Germany}

\author{Tim~J.~Bartley}
\affiliation{Department of Physics, Paderborn University, Warburger Straße 100, 33098 Paderborn, Germany}
\affiliation{Institute for Photonic Quantum Systems (PhoQS), Paderborn University, Warburger Straße 100, 33098 Paderborn, Germany}

\author{Christine~Silberhorn}
\affiliation{Department of Physics, Paderborn University, Warburger Straße 100, 33098 Paderborn, Germany}
\affiliation{Institute for Photonic Quantum Systems (PhoQS), Paderborn University, Warburger Straße 100, 33098 Paderborn, Germany}

\author{Christof~Eigner}
\affiliation{Institute for Photonic Quantum Systems (PhoQS), Paderborn University, Warburger Straße 100, 33098 Paderborn, Germany}

\clearpage

\section{Section S1: Heating curves}
\begin{figure*}[h!]
    \centering
    \includegraphics[width=1\linewidth]{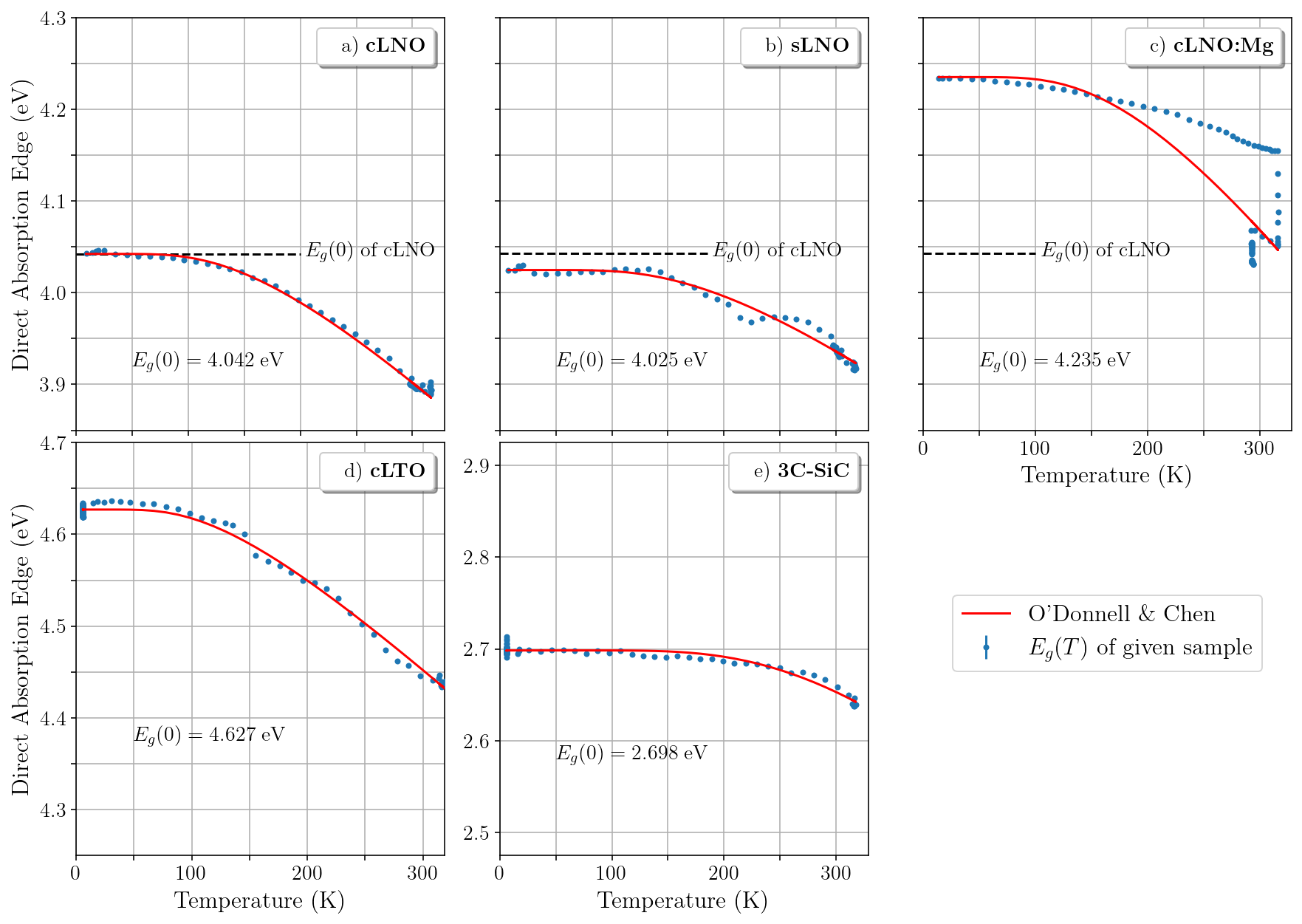}
    \caption{Plots of the direct absorption edge $E_g(T)$ during the heating cycle over a temperature range from approx. 7 K to room temperature for a) cLNO, b) sLNO, c) cLNO:Mg, d) cLTO and e) 3C-SiC fitted with the model of O'Donnell and Chen \cite{odonnell1991}. The band gap at absolute zero of each sample is also shown.}
    \label{fig:placeholder}
\end{figure*}

\begin{table*}[h!]
    \centering
    \begin{tabular}{cccc}
    Material & $ E_g(0\hspace{1mm} \mathrm{K})$ [eV] & $S$ & $\langle\hbar\omega\rangle$ [meV]\\
    \hline
   cLNO & $4.042\pm0.001$ & $7.2\pm0.3$ & $44.2\pm1.9$\\
   sLNO & $4.025\pm0.001$ & $5.8\pm0.6$ & $54.0\pm4.0$\\
   cLNO:Mg & $4.23\pm0.01$ & $10.8\pm4.6$ & $53.6\pm18.0$\\
   cLTO & $4.6270\pm0.0006$ & $7.0\pm0.2$ & $33.7\pm1.4$\\
   3C-SiC & $2.6984\pm0.0004$ & $9.4\pm2.0$ & $96.1\pm7.4$\\
    \end{tabular}
    \caption{Collection of the parameters direct band gap at 0 K, electron-phonon coupling parameter and average phonon energy obtained from the heating curves of the investigated materials.}
\end{table*}
For completeness, sample spectra were also acquired during the heating phase of the cryostat from \~7 K back to room temperature. The resulting bandgaps tend to display some pronounced jumps during the heating process, leading to differences of the parameters $S$ and $\langle\hbar\omega\rangle$ compared with the data acquired during the cooldown phase. These jumps are probably a result of pyroelectric charge build-up. We favoured the cooldown data, because the jumps are less prominent. Nonetheless, the bandgaps at absolute zero are almost identical regardless of the considered data set. Additionally, the electron-phonon coupling parameter of 3C-SiC obtained from the high-temperature data is unusually high; materials with similar crystal structures feature $S$ parameters below 3. Carbon and silicon, for instance, have $S$ parameters of 2.76 and 1.47, respectively \cite{Chiu1998}.

\newpage

\section{Section S2: Influence of Refractive Index \& Thickness}
\begin{figure}[h!]
    \centering
    \includegraphics[width=1\linewidth]{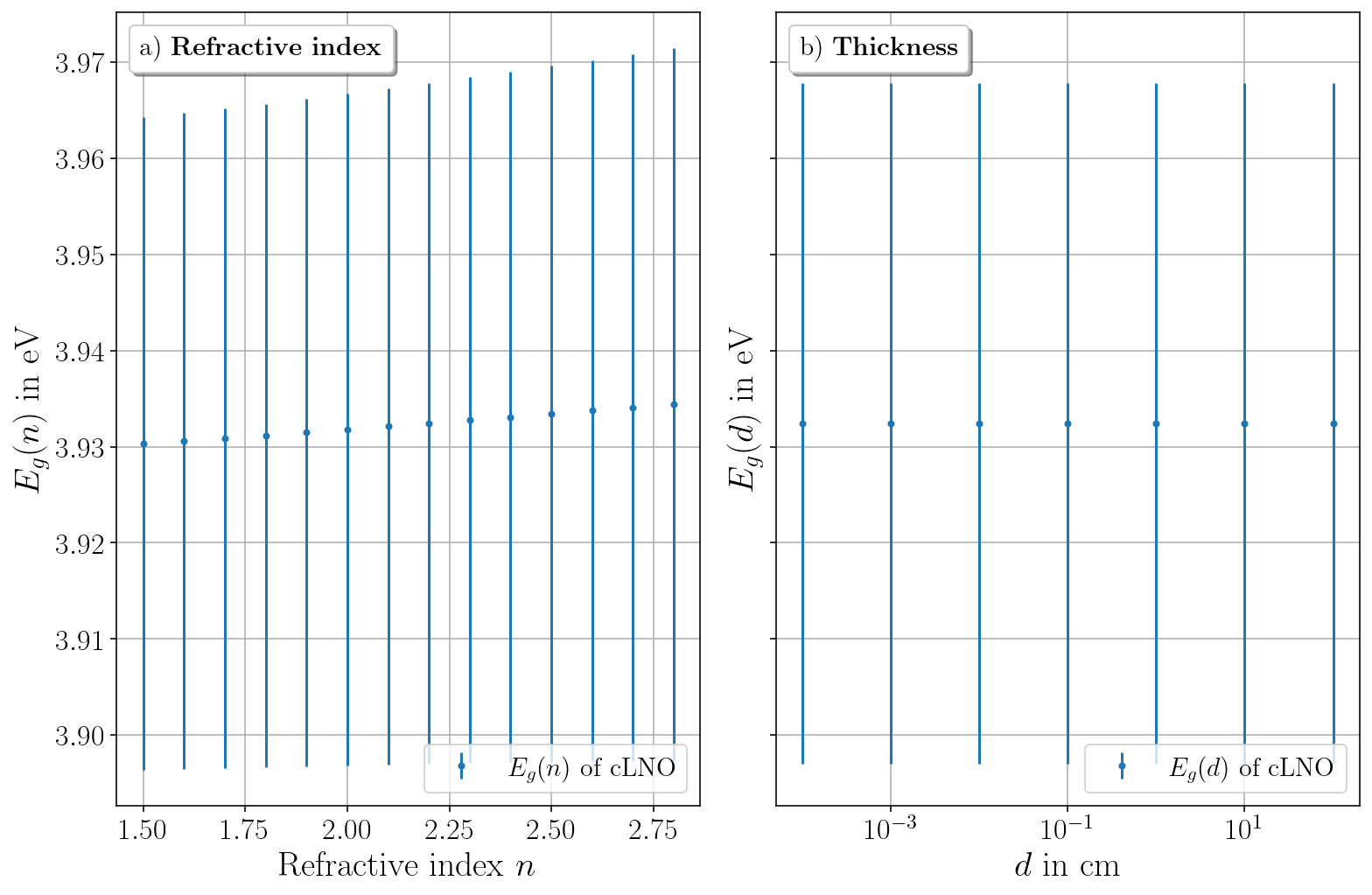}
    \caption{Plot of the influence of different a) refractive indices and b) thicknesses set in PhoQS-Treat on the direct band gap of cLNO at 300 K.}
    \label{fig:dn}
\end{figure}
As indicated in the main text, a constant refractive index and a constant thickness were used for computing the bandgaps due to the complex requirements a Sellmeier equation has to meet. Therefore, multiple validation tests were performed. First, a constant refractive index passed to PhoQS-Treat for the computation of the bandgap of cLNO was varied in a range from 1.5 to 2.8. All other sample parameters and data remained unchanged. While the refractive index clearly impacts the resulting bandgap, the deviations are smaller than the resulting regression uncertainties and therefore negligible.\\
A similar procedure was applied to determine the impact of the thickness. Here, the bandgap is unaffected by the variations. This can be explained with Eq. 3, where the thickness acts merely as a scaling parameter. 

This means that the absolute values of the absorption coefficient $\alpha$ are thickness-dependent but not the x-axis intercept of the Tauc regression or, in other words, the determined bandgap.

\newpage
\section{Section S3: Comparison of Constant Refractive Index and Sellmeier Equations}
\begin{figure}[h!]
    \centering
    \includegraphics[width=1\linewidth]{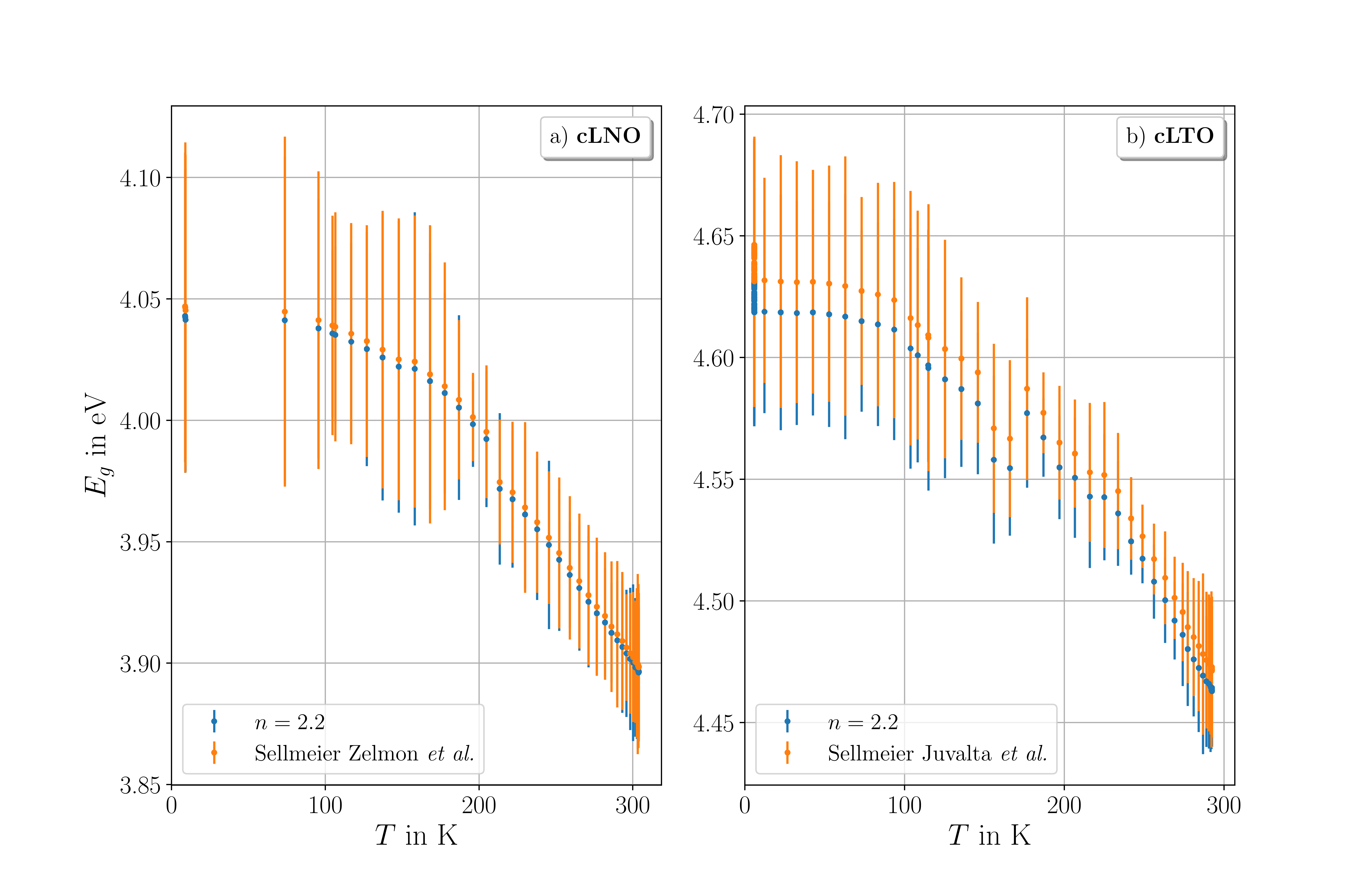}
    \caption{Comparison of the direct band gaps computed with a constant refractive index $n=2.2$ with direct band gaps calculated with the Sellmeier equation of Zelmon \textit{et al.} for a) cLNO and the Sellmeier equation developed by Juvalta \textit{et al.} for b) cLTO. The deviations are in the order of magnitude of approximately 10 meV.}
    \label{fig:zelmon}
\end{figure}
As indicated in the main text, a constant refractive index and thickness were used for computing the bandgaps due to the complex requirements a Sellmeier equation has to meet (polarisation of light, sample stoichiometry, valid for $5~\mathrm{K}\leq T\leq 1000~\mathrm{K}$, applicable below fundamental absorption edge). Therefore, multiple validation tests were performed. Here, the deviations of a constant refractive index compared to a wavelength-dependent one computed with Sellmeier equations are discussed. For cLNO, the equation developed by Zelmon \textit{et al.} is used for comparison while the equation proposed by Juvalta and coworkers is applied for cLTO \cite{zelmon1997, juvalta2006}. The resulting bandgaps deviate by only about 10 meV between constant and wavelength dependent refractive index, with the ones computed with wavelength dependent Sellmeier equations being systematically larger than the ones computed with $n=2.2$. The regression uncertainties are almost unaffected and approximately one order of magnitude larger than the observed deviations, which is why we decided to use a constant refractive index in all computations.

\newpage
\section{Section S4: Derivation of the Expression for the Absorption Coefficient $\alpha$}
We start with the equation of MacFarlane \textit{et al.} \cite{macfarlane1958}:
\begin{equation*}
    \begin{gathered}
        T=\frac{(1-R)^2\exp(-\alpha d)}{1-R^2\exp(-2\alpha d)}\\
        \Leftrightarrow T(1-R^2\exp(-2\alpha d))=(1-R)^2\exp(-\alpha d)\\
        \Leftrightarrow \frac{T-TR^2\exp(-2\alpha d)}{(1-R)^2}=e^{-\alpha d}\\
        \hline
        \mathrm{Substitution:}\\
        x =e^{-\alpha d}; \hspace{0.5cm}x^2=e^{-2\alpha d}\\
        \Rightarrow \alpha=-\frac{1}{d}\ln(x)\\
        \hline\\
        \Rightarrow x= \frac{T-TR^2x^2}{(1-R)^2}\\
        \Leftrightarrow -TR^2x^2-(1-R)^2x+T=0\\
        \Rightarrow x=\frac{(1-R)^2\pm\sqrt{(1-R)^4+4T^2R^2}}{2TR^2}
    \end{gathered}
\end{equation*}
Recalling that $\alpha=-\frac{1}{d}\ln(x)$, the positive solution becomes Eq. 3:
\begin{equation*}
    \alpha = - \frac{1}{d}\ln\biggr(\frac{-(1-R)^2+\sqrt{(1-R)^4+4R^2\cdot T^2}}{2TR^2} \biggl)
\end{equation*}
\newpage
\section{Section S5: Influence of $h\nu$ Factor on the Band Gap}
\begin{figure}[h!]
    \centering
    \includegraphics[width=1\linewidth]{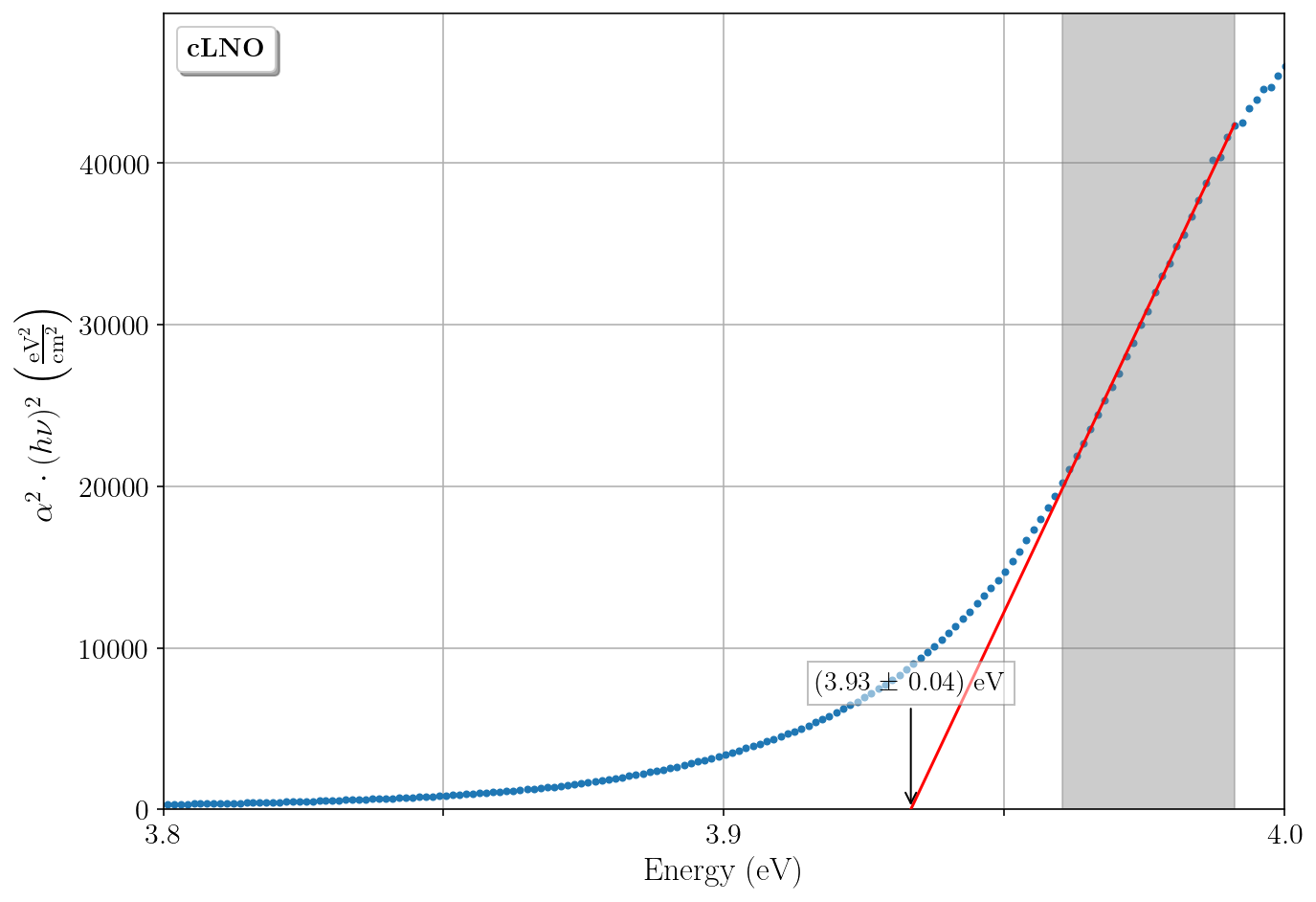}
    \caption{Direct band gap of cLNO taking the $(h\nu)^2$ factor into account. The deviation of the band gap is lower than the accuracy of the measurements.}
    \label{fig:clno_hv}
\end{figure}
According to Eq. 5, the Tauc regression has to be performed over $(\alpha h\nu)^{\frac{1}{\gamma}}$ with $\gamma=\frac{1}{2}$ for a direct electronic transition. However, we decided to perform the regression only over $\alpha^\frac{1}{\gamma}$ because the linear sections important for the bandgap determination are sufficiently small to exclude the $(h\nu)^\frac{1}{\gamma}$ factor. This is illustrated by the plot shown above for the direct bandgap of cLNO taking the factor into account. Neither the absolute value of the determined bandgap nor the corresponding uncertainty deviate from the value computed without the $(h\nu)^2$ factor, at least within the precision of our measurements.\\
On a side note, it is possible to perform the Tauc regression over $(\alpha h\nu)^{\frac{1}{\gamma}}$ as well as over $\alpha^\frac{1}{\gamma}$ using PhoQS-Treat.

\newpage
\section{Section S6: Uncertainty of the High Temperature Data}
\begin{figure}[h!]
    \centering
    \includegraphics[width=1\linewidth]{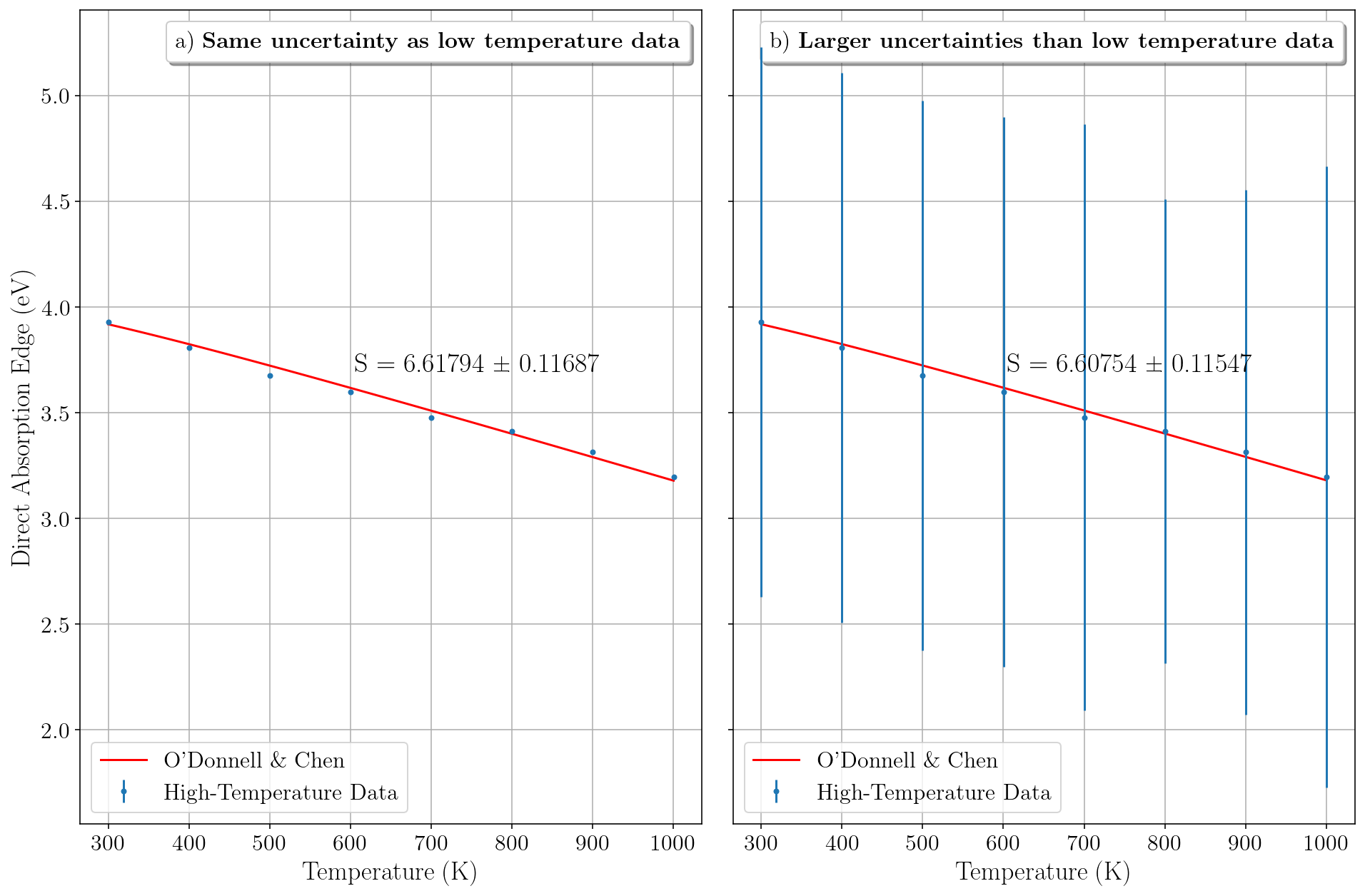}
    \caption{Comparison of the influence of the uncertainty of the high temperature data on the $S$ parameter of cLNO. The deviation between the two values is in the order of magnitude of about 0.2\%.}
    \label{fig:unc_ht}
\end{figure}
As the spectra in the high- and low-temperature regimes were acquired with two different setups, it is important to discuss the conformity of the data. This is especially true concerning the respective uncertainties for the low-temperature data points. Here, the plotted uncertainty is not the mathematical uncertainty of the regression process for each calculated band gap, but rather based on the statistical distribution of the bandgaps. Here, due to the slow cooling process and a long holding time at 7~K, hundreds of spectra at the same temperature could be evaluated. The uncertainty obtained using the standard deviation amounts to 0.0008 eV and is thus two orders of magnitude smaller than the regression uncertainty. In short, the latter is not the relevant uncertainty here as it overestimates the possible range of bandgap values.\\
However, this has not necessarily to be true for the high-temperature data as well. Therefore, we compared the obtained electron-phonon coupling parameters $S$ when using the low-temperature uncertainty of $\pm$0.0008 eV for the high-temperature bandgaps as well and the (actually rather large) regression uncertainties of the high-temperature bandgaps. As becomes evident from Fig. \ref{fig:unc_ht}, the obtained $S$ parameter as well as their uncertainties deviate only insignificantly. Therefore, it is justified to use the small uncertainty of the low-temperature data also for the high-temperature data.

\newpage
\section{Section S7: Data for Fit over whole Temperature Range}
\begin{figure}[h]
    \centering
    \includegraphics[width=1\linewidth]{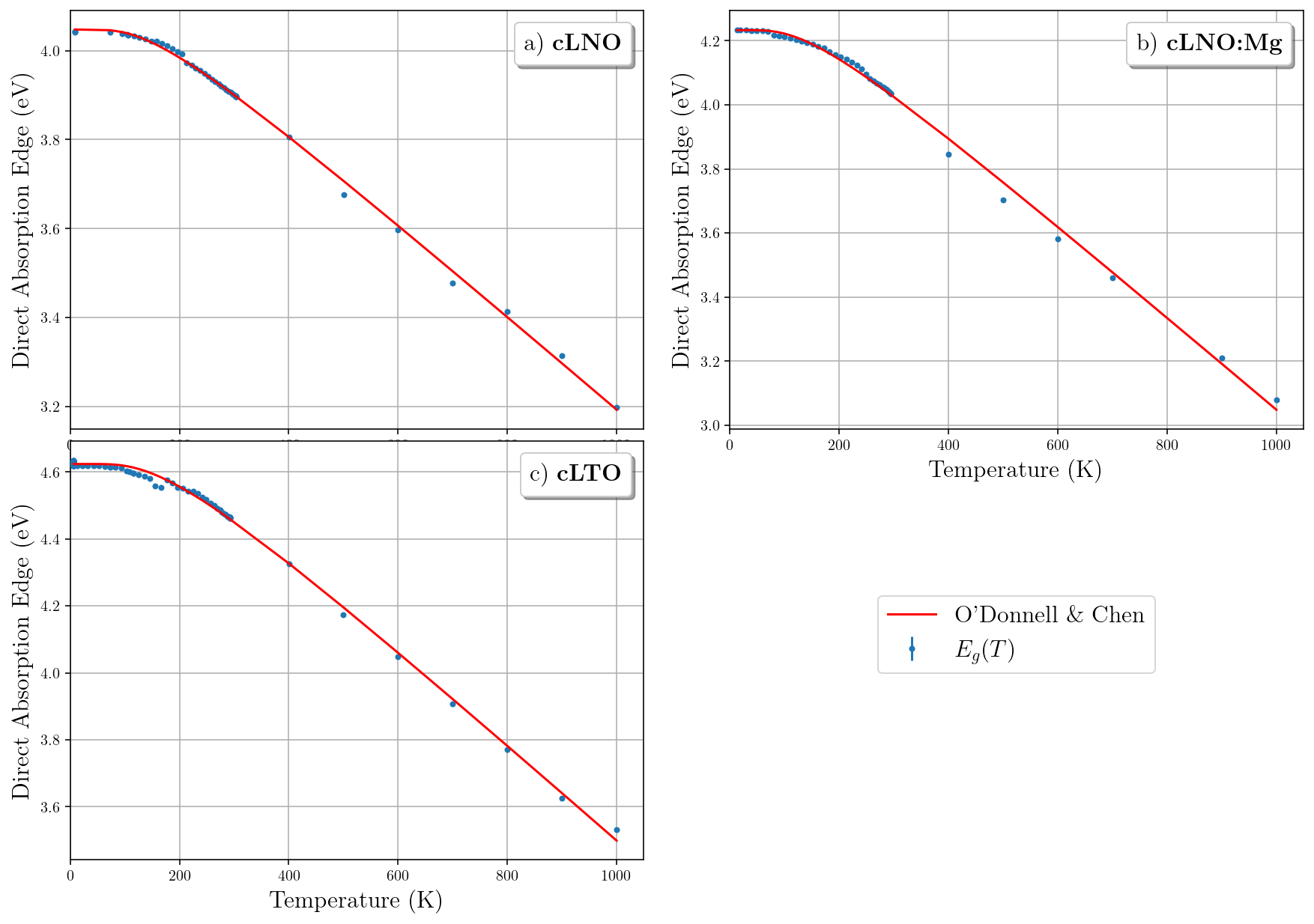}
    \caption{Direct absorption edge data of a) cLNO, b) cLNO:Mg and c) cLTO over a temperature range from approximately 5 K to 1000 K fitted with the model of O'Donnell \& Chen \cite{odonnell1991}.}
    \label{fig:whole_T_range}
\end{figure}

\begin{table}[h]
    \centering
    \begin{tabular}{cccc}
        Material & $E_g(0\hspace{1mm}\mathrm{K})$ [eV] & $S$ & $\langle\hbar\omega\rangle$ [meV]\\
        \hline
        cLNO & $4.048\pm0.003$ & $6.1\pm0.1$ & $35.7\pm1.3$ \\
        cLNO:Mg & $4.233\pm0.005$ & $8.5\pm0.1$ & $34.6\pm1.7$ \\
        cLTO & $4.6240\pm0.0009$ & $8.4\pm0.1$ & $41.9\pm0.7$ \\
    \end{tabular}
    \caption{Collection of the parameters direct band gap at 0 K, electron-phonon coupling parameter and average phonon energy determined with the O'Donnell \& Chen model applied to the band gap data of the investigated materials over the entire considered temperature range.}
    \label{tab:whole_T_data}
\end{table}
The analysis procedure presented in the main text is not the only way to evaluate the data. Instead of analysing the data separately in each temperature regime, which was done to take the different measurement setups into account, it is also possible to fit the model of O'Donnell and Chen over the whole investigated temperature range, leading to slightly different results. For all investigated materials, the bandgap at absolute zero is close to identical compared with the values obtained from individual analysis of the different temperature ranges. However, the $S$ parameter and the average phonon energy are influenced by the (fewer) high-temperature data points by some extent, especially for cLTO. Here, we decided to favour the analysis procedure presented in the main text because the average phonon energy cannot be determined exactly by high-temperature measurements alone for $\langle\hbar\omega\rangle$ defines a tipping point well below room temperature as illustrated by the low-temperature data.

\newpage

\section{Section S8: Data Analysis with PhoQS-Treat}
\begin{figure}[h!]
    \centering
    \includegraphics[width=0.55\linewidth]{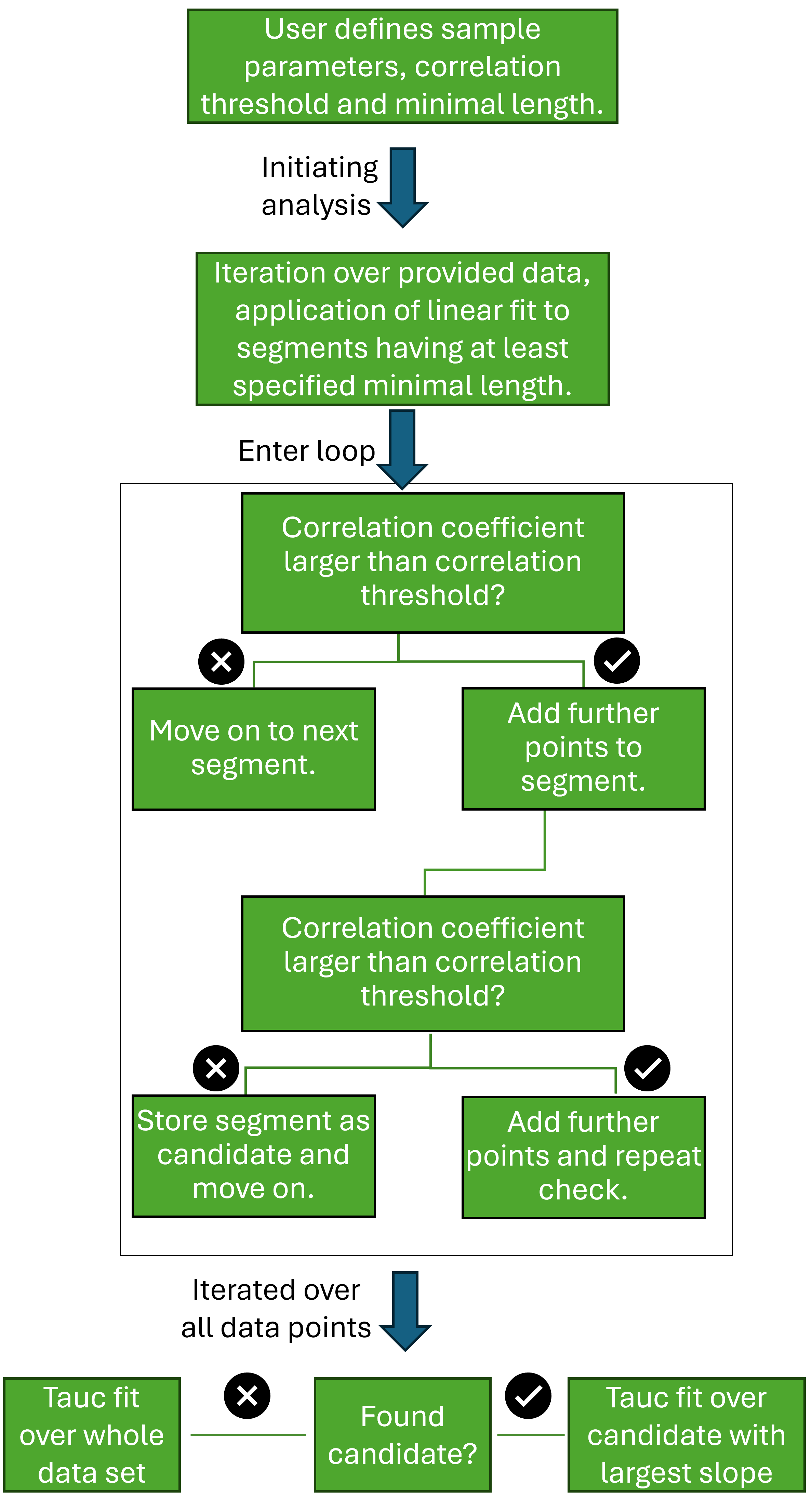}
    \caption{Flowchart illustrating the linear segment detection of PhoQS-Treat.}
    \label{fig:phoqs-treat}
\end{figure}

It is helpful to use appropriate software to accelerate the analysis of large amounts of spectra. This also makes the data analysis process more reproducible.\\
The spectral data were processed using the open-source software PhoQS-Treat, which was developed in this work to provide a reproducible workflow for the analysis of optical transmission spectra. The software allows the user to import reference, dark and sample spectra, optionally smooth the raw spectra using a Savitzky--Golay filter, calculate the transmittance, and extract the absorption coefficient $\alpha$ while accounting for reflection losses at the sample interfaces. The sample thickness $d$ and refractive index $n$ can be specified individually for each sample, enabling batch processing of spectra from samples with different optical parameters. It is also possible to specify dispersion relations for the refractive index $n$. In the present work, a second-order Savitzky--Golay filter with a window length of 80 x-units was applied to the reference and sample spectra prior to further analysis.\\
For the determination of optical band gaps, PhoQS-Treat generates Tauc plots in both wavelength and energy space and therefore performs an automated search for linear regions in the provided data. The algorithm iterates over the data and identifies segments with at least a user-defined minimum length for which the correlation coefficient of a linear fit exceeds a specified threshold, referred to as the correlation threshold. The regression is performed over the whole data set if no segment was able to meet the requirements. In the case that more than one linear segment was found, the algorithm performs a Tauc regression over the segment with the largest slope. This condition is necessary because segments with low transmission below the bandgap behave, to a good approximation, linearly as well. In this work, a threshold of 0.995 was used for all samples, with minimum fit-window lengths of 0.03 eV for the $\alpha^2$ representation and 0.05 eV for the $\sqrt{\alpha}$ representation. In addition, PhoQS-Treat contains a tool for identifying absorption and emission lines based on finding deviations larger than the spectrum’s uncertainty or, as an alternative, on finding the zero crossings in the derivative of the spectrum which are methods from Astropy's specutils package \cite{astropyI,astropyII,astropyIII}. Identified features are fitted with Gaussian, Lorentzian and Voigt profiles, and the resulting fit parameters can be exported for further analysis.

\end{widetext}

\end{document}